\documentclass[fleqn,usenatbib]{mnras}

\usepackage{newtxtext,newtxmath}

\usepackage[T1]{fontenc}

\DeclareRobustCommand{\VAN}[3]{#2}
\let\VANthebibliography\thebibliography
\def\thebibliography{\DeclareRobustCommand{\VAN}[3]{##3}\VANthebibliography}

\usepackage{graphicx}	
\usepackage{amsmath}	
\usepackage[normalem]{ulem}

 \usepackage{comment}

\newcommand{\fgb}{\texttt{fgbuster }}

\title[Synchrotron spectral index and curvature]{Measuring the diffuse Galactic synchrotron spectral index and curvature between 45 and 2300\,MHz}

\author[M. O. Irfan et al.]{
M. O. Irfan,$^{1}$\thanks{E-mail: melis.irfan@ast.cam.ac.uk, giuseppe.puglisi2@unict.it}
G. Puglisi,$^{2,3,4}$
\\
$^{1}$Institute of Astronomy, University of Cambridge, Madingley Road, CB3 0HA, United Kingdom\\
$^{2}$ Dipartimento di Fisica e Astronomia, Universit\`a degli Studi di Catania, via S. Sofia, 64, 95123, Catania, Italy \\
$^{3}$
INAF - Osservatorio Astrofisico di Catania, via S. Sofia 78, 95123 Catania, Italy\\
$^{4}$
INFN - Sezione di Catania, Via S. Sofia 64, 95123 Catania, Italy\\
}

\date{Accepted XXX. Received YYY; in original form ZZZ}

\pubyear{\the\year{}}

\begin{document}
\label{firstpage}
\pagerange{\pageref{firstpage}--\pageref{lastpage}}
\maketitle

\begin{abstract}
We present an all-sky map of the synchrotron spectral index and curvature between 45 and 2300\,MHz at a resolution of 1$^{\circ}$ calculated from a combination of numerous partial sky empirical measurements. We employ a least-squares parametric fit which relies on removing a free-free emission template and a component separation technique which fits for both synchrotron and free-free emission.  We compare  our diffuse sky model estimates against those derived from the models widely used in the community (e.g. \texttt{pysm3} and \texttt{GSM})  employing  external data sets that were not included in the  estimation process. Our evaluation focuses on identifying the enhanced consistency at both the map level and in pixel-to-pixel correlations, allowing for a more robust verification of our model's performance. We find our parametric, least-squares synchrotron estimate to be the most reliable across radio frequencies as it consistently provides sky models with average accuracies (when compared to empirical data) of around 20 per cent, whilst other model performances range on average between 10 and 70 per cent accurate.  The results obtained have been made publicly accessible online and can be utilized to further develop and refine models of Galactic synchrotron emission.
\end{abstract}




\section{Introduction}
Diffuse Galactic synchrotron emission is caused by charged, relativistic particles, such as cosmic-ray electrons, propagating along the field lines of our Galactic magnetic field \citep{strong}. It is the dominant diffuse Galactic emission at frequencies under 1\,GHz across the majority of the sky, excluding the central Galactic plane in both intensity and polarization \citep{bennett}. Synchrotron emission is of particular interest to several communities due to its predominance in a very wide range of frequencies and because of its  complex spectral and spatial behavior. 
On one hand, for the interstellar medium community measurements of diffuse synchrotron emission place constraints on the Galactic magnetic field strength as well as cosmic-ray electron propagation \citep{pad21, bracco}, provide an understanding of the diffuse background required for transient and supernova remnant detection \citep{trans, erosita} and probe exotic physics such as dark matter annihilation \citep{dm} and outflows/super bubbles from the Galactic center \citep{haze}. Within the cosmological community diffuse Galactic synchrotron emission provides a pernicious foreground for both  21\,cm intensity mapping and the Cosmic Microwave Background (CMB).

21\,cm experiments aim to measure the large angular scale, unresolved, integrated emission from neutral hydrogen atoms across redshift thus mapping the formation of Large Scale Structure in the Universe and determining the redshift of the Epoch of Reionisation. 21\,cm experiments are either global experiments which aim to measure the average neutral hydrogen temperature (e.g. EDGES \citep{edges}, LEDA \citep{leda}, REACH \citep{reach}, SARAS \citep{saras}) or intensity mapping experiments which measure the anisotropies around this average (e.g. BINGO \citep{bingo}, CHIME \citep{chime}, HIRAX \citep{hirax}, MeerKLASS \citep{meerklass}). Global and intensity mapping 21\,cm experiments measure within the 50-1420\,MHz frequency range and so suffer contamination from synchrotron emission in intensity and also in polarization through the so-called \emph{Polarization-to-Intensity } leakage of Stokes parameters Q and U to I \citep{shaw14}. 
Similarly, the constraints of cosmological parameters as well as the detection of the faint  $B-$mode polarization signal from  CMB data is strongly contaminated by Galactic foregrounds, if not properly taken into account and/or removed from the CMB intensity and polarization maps \citep{so,litebird,planckPS}.

 The synchrotron emission spectrum at each pixel on the sky ($p$) is usually modeled as a power law,
\begin{equation}
\label{eq:ind}
T_{\nu}(\nu, p) = T_{\nu_{0}}(\nu, p) \left(\frac{\nu}{\nu_{0}} \right)^{\beta_\text{s}(p)},
\end{equation}
where the synchrotron spectral index $\beta_{\rm s}$ is measured between the frequencies $\nu$ and $\nu_0$. $T_{\nu_{0}}$ is a synchrotron temperature map at a particular frequency. As synchrotron emission decreases with increasing frequency, the lowest frequency, highest resolution all-sky radio map available is often considered to be the best proxy for a synchrotron emission amplitude template. Since its inception, in the early 1980s, the Haslam 408\,MHz \citep{haslam} all-sky map at a resolution of 56 arcmin has universally been used to represent a full sky view of synchrotron emission.   
It is well understood that the synchrotron spectral index should change within different lines of sight  in the sky, due to energy losses as the charged relativistic particles travel through the Galaxy. A widely accepted measurement of the synchrotron spectral index across the full sky was made in \cite{mamd}, where the authors explored several methods to very different effect. All the methods involve trying to form an all-sky map at 5$^{\circ}$ resolution between 408\,MHz and 23\,GHz using Haslam and {\emph{WMAP}} \citep{wmap} data. 
The \citet{mamd} spectral index has been vastly employed, particularly in modeling synchrotron emission in the frequency regime of typical CMB experiments (20-100 GHz). In fact, the latest models of the Python Sky Model \citep{pysmOrg} \texttt{pysm3} suite, \citet{pysm3}, employ  the \citet{mamd} template together with the recently available S-PASS data \citep{Krachmalnicoff_2018}. Moreover, the limit in resolution has been overcome by artificially adding  small scales in the spectral index map \citep{pysm3}.

Not only is the spectral index expected to change spatially, it is also believed to vary across frequency. Throughout the early 2000s an increasing number of researchers were noticing a discrepancy between the spectral index measured at radio frequencies by ground-based telescopes and microwave measurements made by satellites. \citet{deOliveira2008} note that the spectrum of synchrotron emission seems to steepen with increasing frequency after a consideration of over twenty partial sky maps between 10\,MHz and 100\,GHz. The ARCADE2 balloon-borne experiment \citep{arcade} measured a region of the sky between 3 and 90\,GHz and fit an empirical form for the believed change in spectral index over frequency, denoted as spectral index `curvature ($c_{s}$)': 
\begin{equation}
\label{eq:curve}
\beta_{\text{s}}(p) = \beta_{\text{s},0}(p) + c_{s} \, {\rm{ln}} \left( \frac{\nu}{\nu_{0}} \right),
\end{equation}
where they measured $\beta_{s, 0}(p)$  to be $-2.60\pm 0.04$ and $c_{s}=-0.052 \pm 0.005$ , respectively at $\nu_{0} = 310$\,MHz \citep{kogut}. The 21\,cm experiments EDGES \citep{edgesSY} and MeerKLASS \citep{minebeta} have both placed constraints on the synchrotron spectral index curvature and these were found to agree with each other and to the ARCADE2 measurement within 1$\sigma$. Recently, \citet{q_cs} used simulated data from the \texttt{PySM} at QUIJOTE-MFI2, {\emph{WMAP}} and {\emph{Planck}} frequencies to forecast constraints on the per-pixel spectral index across the northern hemisphere at 1$^{\circ}$ resolution. The synchrotron spectral index was predicted to range between $-3.4$ and $-2.7$, when curvature was not considered. However, at frequencies above 10\,GHz the authors noted that the synchrotron emission magnitude would be too weak to place strong constraints on a curved synchrotron emission model. Previous work using empirical MFI1 polarized intensity data alongside polarized intensity {\emph{WMAP}} and {\emph{Planck}} data \citep{2023q} tested for both a simple power law model and spectral index curvature and found no significant motivation for curvature, while \citet{q_steep} note that lower frequency data would need to be included alongside QUIJOTE data to robustly constrain curvature.

Simulations of diffuse Galactic emissions have particular use within the 21\,cm and CMB communities as both these communities aim to detect a faint background signal amidst contamination from bright Galactic foregrounds. Both foreground removal and foreground avoidance require a detailed knowledge of the emission spectral and spatial behavior. Additionally, instrumental calibration schemes may require an initial estimate of the radio sky temperature and so an understanding of synchrotron emission both spatially and spectrally is required. Total diffuse emission models, such as the Global Sky Model (\texttt{GSM} \citep{gsm}), \texttt{pysm3} and Bayesian Global Sky Model \citep{bayes}, are especially useful as they provide temperature estimates for the combined diffuse Galactic emission temperature (which will be some combination of synchrotron, free-free, anomalous microwave and thermal dust emission) per pixel across a large frequency range for the whole sky.

Lately, the community has been focusing on improving the synchrotron models \citep{cnnBeta, synax, tigress}. This involves accurately accounting for the emission spatial and spectral variations (e.g. by incorporating the curvature parameter), and by overcoming the limitation of low angular resolution synchrotron templates. The objective of this study is to tackle both these two goals by combining information from a total of 37 distinct partial and all-sky maps, each of which are available at varying resolutions. This integration is aimed at generating comprehensive maps that represent an updated template of the diffuse Galactic synchrotron spectral index and curvature intended to be  employed across a wide range of frequencies, specifically from 45 to 2300\,MHz.

The paper is organized as follows: in \autoref{sec:data} we introduce the publicly available data and \autoref{sec:methods} details the processing steps required to combine the data sets into a parametric fit of the spectral index and curvature. We explore two ways to perform the fit, a least-squares parametric fit which relies on  removing a free-free emission template and a component separation technique which fits for both synchrotron and free-free emission. In \autoref{sec:results} we form an all-sky model of synchrotron emission by scaling the Haslam data using the fitted spectral index and curvature maps. 
Lastly, in \autoref{sec:conc} we present our conclusions.   

\section{The data}
\label{sec:data}
In this work we make use of publicly available radio surveys that cover both the northern and southern hemispheres. \autoref{table:dat1} summarizes the main observational details of the publicly available surveys employed within this work: Maipu/MU \citep{mu}, LWA1 \citep{lwa}, OVRO-LWA \citep{ovro}, Landecker \citep{land}, Haslam \citep{haslam}, GMIMS-HBN \citep{hbn} and GMIMS-STAPS \citep{staps}. For the 408\,MHz data we make use of the destriped map reprocessed by \citet{rem}. We remark that the following analysis is performed without applying \emph{color-corrections} on the whole data set \footnote{This effect, albeit smaller than the ones accounted for in \autoref{sec:zero}, could effectively  bias the estimates of spectral parameters. However, for several surveys, the color-correction factors have not been reported in the literature; we therefore decide to  neglect them.}.

\begin{table}
\begin{center}
\addtolength{\tabcolsep}{-0.55em}
\begin{tabular}{||c c c c c c||} 
 \hline
Survey & Freq. (MHz) & Res. ($^{\circ}$) & $\sigma_{{\rm{cal}}}$ ($\%$) & $\sigma_{{\rm{t}}}$ & Coverage \\ [0.5ex] 
 \hline\hline
 Maipu/MU  & 45 & 5.0 & 7.0 & 0.23$^{b}$ &Full\\
  \hline
   OVRO-LWA & 41.8 & 0.29 & 5.0 & 0.8$^{a}$& North \\
 \hline
 OVRO-LWA & 47.0 & 0.27 & 5.0 &0.8$^{a}$ & North  \\
 \hline
 OVRO-LWA & 52.2 & 0.25 & 5.0 &0.8$^{a}$ & North  \\
 \hline
  OVRO-LWA & 57.5 & 0.25 & 5.0 &0.8$^{a}$ & North  \\
   \hline
   OVRO-LWA & 62.7 & 0.25 & 5.0 &0.8$^{a}$ & North \\ 
 \hline
 OVRO-LWA & 67.9 & 0.25 & 5.0 &0.8$^{a}$ & North  \\
 \hline
 OVRO-LWA & 73.2 & 0.25 & 5.0 &0.8$^{a}$ & North  \\
 \hline
   LWA & 45 & 3.6 & 5.0 & 20$^{a}$ &North \\
   \hline
  LWA1 & 50 & 3.3 & 5.0 & 16$^{a}$ &North  \\
   \hline
   LWA1 & 60 & 2.7 & 5.0 &10$^{a}$ & North\\ 
 \hline
 LWA1 & 70 & 2.3 & 5.0 & 7$^{a}$ & North \\
 \hline
 LWA1 & 74 & 2.2 & 5.0 &6$^{a}$ & North  \\
 \hline
 LWA1 & 80 & 2.0 & 5.0 & 5$^{a}$ & North  \\
  \hline
Landecker  & 150 & 5.0 & 5.0 & 0.62$^{b}$ & Full  \\
 \hline
  Haslam & 408 & 0.93 & 10.0 & 0.1$^{b}$  & Full \\
 \hline
GMIMS-HBN & 1383 & 0.67 & 8.0 & 0.02$^{b}$& North \\
 \hline
GMIMS-HBN & 1418 & 0.67 & 8.0 & 0.02$^{b}$ & North \\
\hline
GMIMS-HBN & 1456 & 0.67 & 8.0 & 0.02$^{b}$ & North \\
\hline
GMIMS-HBN & 1487 & 0.67 & 8.0 & 0.02$^{b}$ & North \\
\hline
GMIMS-HBN & 1499 & 0.67 & 8.0 &0.02$^{b}$  & North \\
\hline
GMIMS-HBN & 1521 & 0.67 & 8.0 & 0.02$^{b}$ & North \\
\hline
GMIMS-HBN & 1614 & 0.67 & 8.0 & 0.02$^{b}$ & North \\
\hline
GMIMS-HBN & 1625 & 0.67 & 8.0 & 0.02$^{b}$ & North \\
\hline
GMIMS-HBN & 1660 & 0.67 & 8.0 & 0.02$^{b}$ & North \\
\hline
GMIMS-HBN & 1700 & 0.67 & 8.0 & 0.02$^{b}$ & North \\
\hline
GMIMS-HBN & 1712 & 0.67 & 8.0 & 0.02$^{b}$ & North \\
\hline
GMIMS-STAPS & 1324 & 0.33 & 10.0 & 0.016$^{b}$& South \\
\hline
GMIMS-STAPS & 1349 & 0.33 & 10.0 & 0.016$^{b}$& South \\
\hline
GMIMS-STAPS & 1374 & 0.33 & 10.0 & 0.015$^{b}$& South \\
\hline
GMIMS-STAPS & 1456 & 0.33 & 10.0 & 0.013$^{b}$& South \\
\hline
GMIMS-STAPS & 1524 & 0.33 & 10.0 & 0.012$^{b}$& South \\
\hline
GMIMS-STAPS & 1609 & 0.33 & 10.0 & 0.011$^{b}$& South \\
\hline
GMIMS-STAPS & 1628 & 0.33 & 10.0 & 0.010$^{b}$& South \\
\hline
GMIMS-STAPS & 1700 & 0.33 & 10.0 & 0.009$^{b}$& South \\
\hline
GMIMS-STAPS & 1749 & 0.33 & 10.0 & 0.008$^{b}$& South \\
\hline
GMIMS-STAPS & 1770 & 0.33 & 10.0 & 0.008$^{b}$& South \\
\hline
  \hline
\end{tabular}
\caption{The data sets employed in this work; survey name, central frequency, angular resolution, calibration errors, instrumental noise level and sky coverage. Instrumental noise ($\sigma_{{\rm{t}}}$) is given either in Jy per beam ($^{a}$) or K ($b$).}
 \label{table:dat1}
\end{center}
\end{table}

\section{Methods}
\label{sec:methods}

Total emission ($T_{\rm{tot}}$) measurements of the sky (per pixel $p$) across radio frequencies ($\nu$) contain several temperature contributions:

\begin{equation}
\begin{aligned}
\label{eq:basic}
T_{\rm{tot}}(p, \nu) &= T_{\rm{diff}}(p, \nu) + T_{\rm{ps}}(p, \nu) + T_{\Delta\rm{cosmo}}(p, \nu) + T_{\rm{cosmo}}(\nu) \\ &+ T_{\rm{inst}}(\nu),
\end{aligned}
\end{equation}
where $T_{\rm{diff}}$ is the contribution from diffuse Galactic emissions, $T_{\rm{ps}}$ is from point sources large enough to be measured as distinct features by the instrumental beam and $T_{\rm{cosmo}}$ is from cosmological backgrounds such as the CMB, the neutral hydrogen background (21\,cm emission) and the unresolved radio point sources collectively known as the Cosmic Radio Background (CRB). Cosmological backgrounds contribute a constant single value across all pixels of the sky, such as the CMB monopole of 2.725\,K, as well as cosmological anisotropies ($T_{\Delta\rm{cosmo}}$) around this monopole value which vary across pixels. $T_{\rm{inst}}$ represents a frequency dependent monopole inherent to each individual experiment. Whilst, all of the observational data used in this work are calibrated from receiver units into Kelvin, only two surveys have absolutely calibrated temperature zero-levels: Haslam 408\,MHz and Landecker 150\,MHz. This means that all other surveys are measuring temperature variations across the sky above an arbitrary zero-level determined by their own data reduction pipeline. 

In order to measure the diffuse Galactic synchrotron emission spectral index, we must isolate the diffuse Galactic synchrotron emission at each pixel and frequency. For this we make two main assumptions about radio data: 
\begin{enumerate}
    \item Diffuse Galactic emission at frequencies under 3\,GHz consists of only synchrotron and free-free contributions. Anomalous microwave and thermal dust emissions are negligible within this frequency range.
    \item  The CMB, CRB and 21\,cm spatially varying signals at each frequency are so small in comparison to the spatially varying signal coming from our own Galaxy at each frequency that they can be considered negligible.  
\end{enumerate}

Rewriting Eq.~\ref{eq:basic} with these assumptions in mind yields: 

\begin{equation}
\label{eq:new}
T_{\rm{tot}}(p, \nu) = T_{\rm{sync}}(p, \nu) + T_{\rm{ff}}(p, \nu) + T_{\rm{ps}}(p, \nu) + T_{\rm{back}}(\nu),
\end{equation}
where $T_{\rm{sync}}(p, \nu)$ and $T_{\rm{ff}}(p, \nu)$ represent diffuse Galactic synchrotron and free-free emission, respectively and $T_{\rm{back}}(\nu)$ is the combination of any monopole temperatures present. All that is now required to fit a synchrotron spectral index across the different surveys is to 1) calculate and subtract the frequency-dependent zero levels, 2) model and remove the diffuse free-free emission and 3) mask out and inpaint the point source contributions.    

\subsection{Zero-Level calibration}
\label{sec:zero}
In general fewer surveys with absolute zero-level calibration exist than those without, as an additional reference black-body emitter is required within the instrumentation to provide regular zero-level checks. However, it is possible to measure a survey zero-level through comparison with another, zero-level calibrated survey \citep{jonas}.

Synchrotron emission can be modeled as a power law. If an observation within a single pixel across two frequencies represents pure synchrotron emission then \autoref{eq:ind} holds true. The total temperature measured will include the survey zero-levels:
\begin{equation}
T_{\rm{sync}}(p, \nu_{2}) + c_{v2} = \left (T_{\rm{sync}}(p, \nu_{1}) - c_{v1} \right)\times \left(\frac{\nu_{2}}{\nu_{1}} \right )^{\beta_{\rm sync}}
\end{equation}

The Haslam 408\,MHz all-sky survey has both a temperature scale and zero-level calibration and, thanks to its widespread use, the monopole due to the combined CMB, CRB and 21\,cm emission has been calculated as $8.9 \pm 1.3$\,K \citep{wehus}. Therefore for any survey compared with the Haslam survey:

\begin{equation}
T_{\rm{sync}}(p, \nu_{2}) = (T_{\rm{tot}} - 8.9)(p, 408) \times \left(\frac{\nu_{2}}{408} \right )^{\beta_{\rm sync}} - c_{v2}
\end{equation}
As long as the maps at both frequencies are observing the same emission then the above equation can be fit using linear regression providing a value for $c_{v2}$ as the fitted y-intercept. If, however, the difference in frequency means that one survey only observes synchrotron emission whilst the seconds survey observes synchrotron and non-negligible free-free emission then the two data sets will have a poor correlation coefficient and the linear regression will fail.  

The zero-level value for each frequency map was calculated by: 

\begin{enumerate}
    \item Smoothing all maps to a common resolution assuming Gaussian beams.
    \item Each map for a given frequency $\nu$ was subdivided into sub-regions corresponding to HEALPix pixels at $N_{{\rm{side}}} =8 $ \citep{gorski}. Therefore each sub-region encodes 1024 pixels out of the original map which is at N$_{\rm{side}}$ = 256.
    \item Linear regression was performed within each sub-region between the considered frequency $\nu$ and 408\,MHz after having removed the 8.9\,K offset from the Haslam map.
    \item The mean of all the region offsets was used as the $\nu$ zero-level, with the standard error on the mean providing the offset error.  
\end{enumerate}

The errors associated with each data set are just the survey calibration errors except for the Haslam data where the total error ($\sigma_{\rm{haslam}}$) is calculated as: 

\begin{equation}
\sigma_{\rm{haslam}} = \sqrt{{\sigma_{\rm{cal}}}^2 + {\sigma_{\rm{offset}}}^2},
\end{equation}
where ${\sigma_{\rm{offset}}}$ is 1.3\,K. We choose to split the sky into sub-regions to provide multiple attempts to measure the single map offset value as we expect our fit to start to fail within regions where there is a non-negligible contribution of free-free emission. As free-free and synchrotron emission follow different spectral behaviors a temperature-temperature plot across two different frequencies for one patch of the sky will not produce strongly linearly correlated results.

\autoref{fig:ohists} and \autoref{fig:omaps} display the offset values calculated within each sub-region for the single-dish empirical data at each frequency as both maps and a histogram distributions. The mean offset was used as the final map offset, with the standard error on the mean providing the uncertainty. Note that we do not calculate the zero-level offset for the OVRO-LWA data as they are interferometric data and therefore missing any angular scales larger than the shortest array baselines can measure. The calculated offsets for each single-dish observation are listed in \autoref{tab:offvals}. We obviously do not need to calculate an offset for the Haslam data but we do calculate one for the Landecker 150\,MHz survey despite the fact that it also has an absolute zero-level. This is done because we still want to remove the combined CMB and CRB temperature from the map and we only have an understanding of the CMB monopole at 2.725\,K. 

\subsection{Removal of free-free template and least-squares fitting}
\label{sec:ff}
After the offset removal, each map contains some mixture of both diffuse free-free and diffuse synchrotron emission, as well as the point source emission contribution. In order to isolate pure diffuse synchrotron emission we must either remove a free-free emission template or specifically fit for the synchrotron emission. We explored both these options; this section describes the first approach which removes a free-free template and then performs a least-squares fit to the remaining emission (assumed to be pure synchrotron).  

\begin{figure}
 \centering
  {\includegraphics[width=0.8\linewidth]{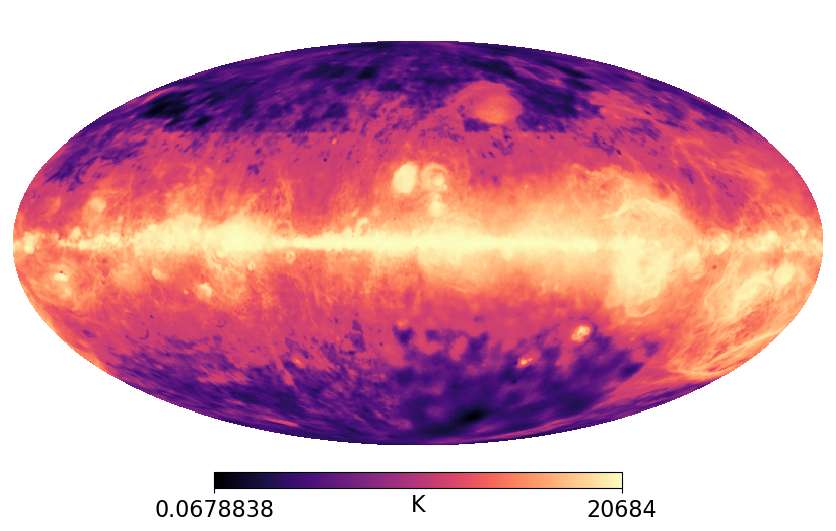}}\\
  {\includegraphics[width=0.8\linewidth]{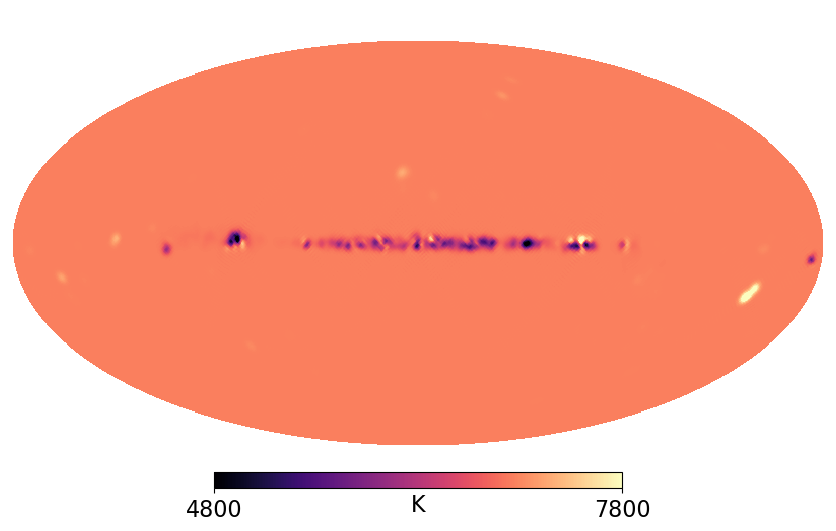}}\\
 \caption{{\emph{Top:}} Emission measure template used in this work. {\emph{Bottom:}} Electron temperature template used in this work.}
 \label{fig:ff}
   \end{figure} 
   
Free-free emission can be parametrized using electron temperature ($T_{e}$) and emission measure (EM) as follows:  

\begin{equation}
T_{{\rm{ff}}} = T_{e}\,(1-e^{-\tau}),
\end{equation}
where
\begin{equation}
\tau = 0.05468 \, T_\mathrm{e}^{-3/2} \, \nu_{9}^{-2}  \,{\rm{EM}} \, g_{\rm ff},
\end{equation}
where
\begin{equation}
 g_{\rm ff}(\nu, T_\mathrm{e}) = \ln \left\{ \exp \left[5.960 - \frac{\sqrt{3}}{\pi} \ln(\nu_{\rm GHz}  T_{4}^{-3/2})  \right] + e \right\}, 
\end{equation}
and $\nu_{9}$ and $T_{4}$ are $\nu ={10^{-9}}$\,Hz and $T_\mathrm{e}= 10^{-4}$\,K respectively. 

\begin{figure}
 \centering
  {\includegraphics[width=0.8\linewidth]{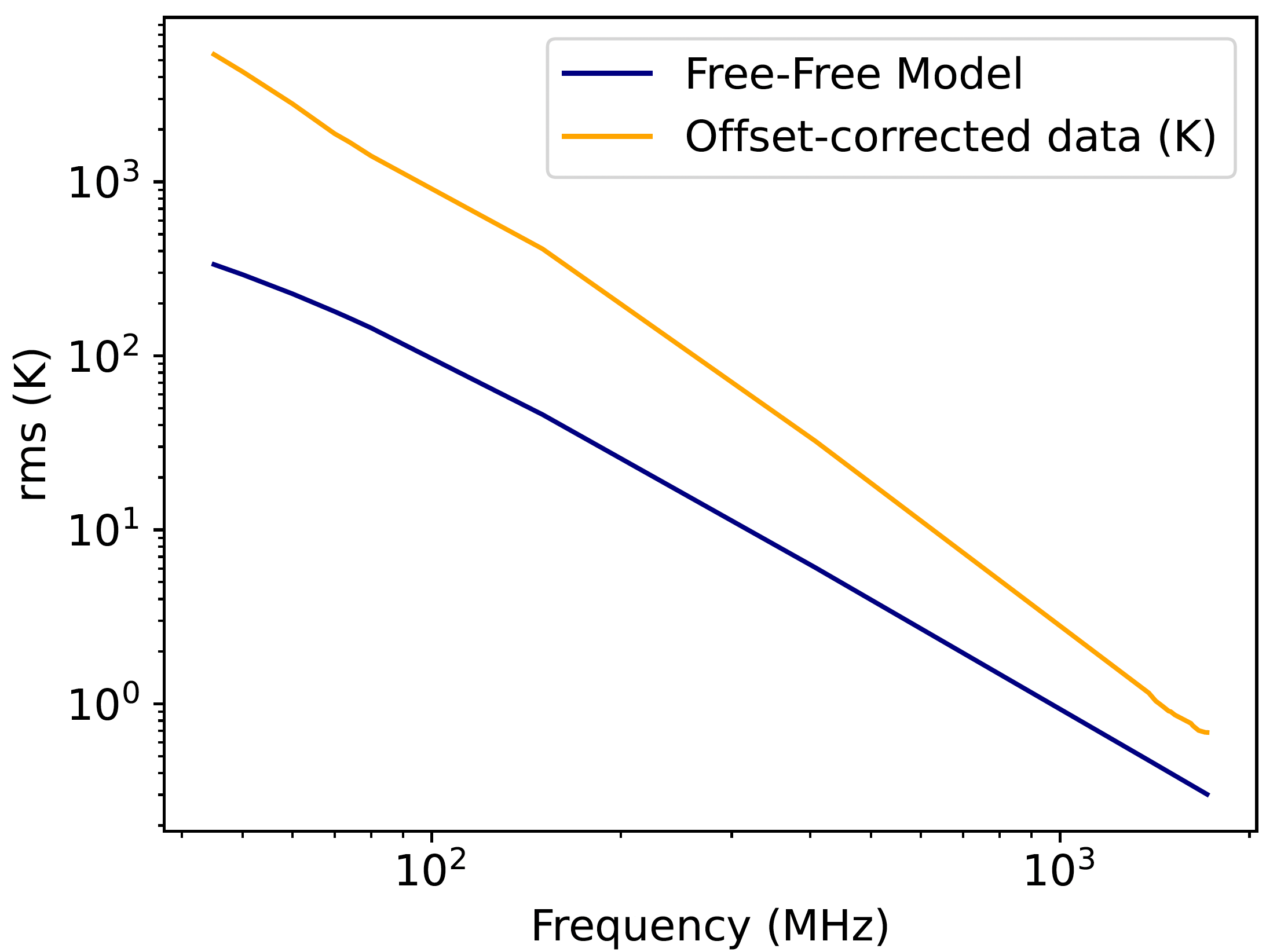}}
 \caption{RMS of maps as a function of frequencies estimated in the common footprint for the northern hemisphere data at N$_{\rm{side}}$ = 256 all smoothed to a common 5$^{\circ}$ resolution.}
 \label{fig:rms}
   \end{figure}

\autoref{fig:ff} presents the emission measure template and electron temperature template used in this work to simulate free-free emission at each total emission map frequency. We employ the \citet{huts} EM map, obtained with a joint inference model of the {\emph{Planck}} \citep{ffparam} and H-$\alpha$ \citep{draine2011, fink} emission measure maps. \citet{huts}  publicly released the EM map  at a 6 arcmin resolution and N$_{{\rm{side}}}$ = 256. The $T_e$ map we employ was produced through the application of the Bayesian component separation framework Commander to the {\emph{Planck}} PR2 data \citep{ffparam}, available at a resolution of $1^{\circ}$ and N$_{{\rm{side}}}$ = 256. \autoref{fig:rms} illustrates the typical contribution of free-free emission (from our model) to the total diffuse Galactic emission by plotting the temperature RMS across the single dish empirical observations smoothed to 5$^{\circ}$ resolution.  

 \begin{figure}
 \centering
  {\includegraphics[width=0.8\linewidth]{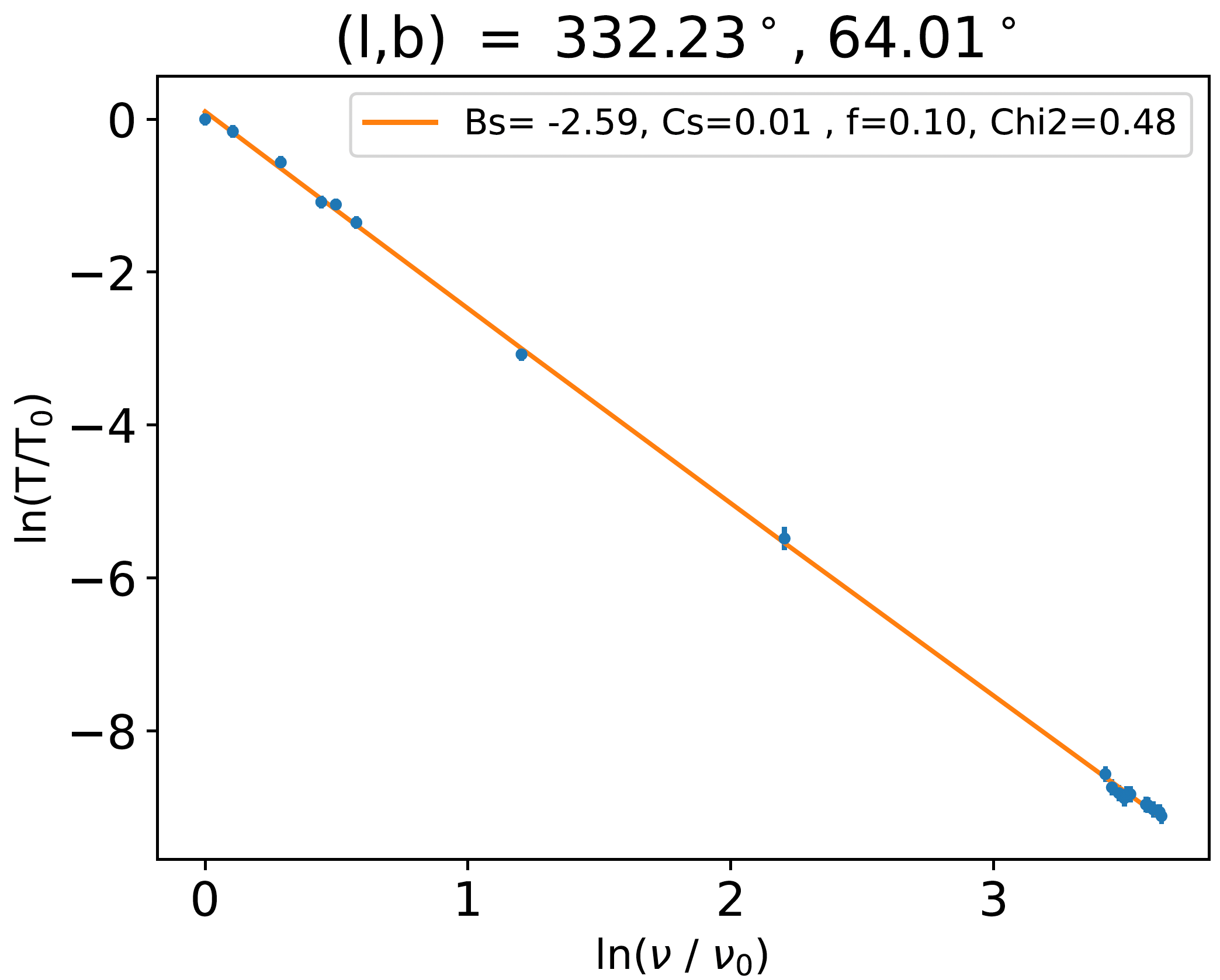}}
 \caption{An example of the per-pixel fit (Galactic latitude and longitude noted on the figure) performed to determine the synchrotron spectral index and magnitude of spectral index curvature. Empirical data points are given in blue; the best fit to the data is the orange line.}
 \label{fig:fitex}
   \end{figure} 
   
After the offset and free-free emission removal the empirical maps contain only diffuse Galactic synchrotron and point source emissions. We choose to leave the point sources in the maps for the joint fit and to only remove them in the final fitted parameter maps. The logic behind this was driven by the fact that one map viewing the sky at a resolution of 56 arcmin will see the point sources smoothed with a 56 arcmin FWHM beam thus spreading some of the point source temperature into the diffuse emission temperature. When we smooth a second, finer resolution map to 56 arcmin to compare with the first map we also want the point source temperature contribution to be smoothed into the diffuse temperature contribution as that is the most faithful representation of the sky as measured through a 56 arcmin beam. Therefore we choose to ignore the points sources for the fit in favor of masking them out in the final parameter maps. 

Assuming pure diffuse Galactic synchrotron emission the fitted model becomes:

\begin{equation}
\frac{T_{\rm{sync}}(p, \nu)}{T_{\rm{sync}}(p, \nu_{0})}  =  \left(\frac{\nu}{\nu_{0}} \right )^{\beta_{\text{s}}},
\end{equation}
where $\beta_{\text{s}}$ is given by \autoref{eq:curve} and the two free parameters are the spectral index ({$\beta_{\text{s}}$) and curvature ($c$). We choose $\nu_{0}$ to be the lowest frequency available from the empirical data as the synchrotron emission in this map has the highest signal to noise ratio due to the negative spectral index of synchrotron emission.

A per-pixel fit of empirical data can only be performed when each pixel represents the same angular information on the sky so each empirical map must be smoothed to a common resolution and downgraded to the same N$_{{\rm{side}}}$. We perform four fits across four different data combinations. This is necessary as several of the empirical maps are northern or southern hemisphere only, some data sets are single-dish and have a resolution limit and other data sets are interferometric and are missing large scale information on the sky. The four combinations are:

\begin{enumerate}
    \item {\bf{Northern coarse}} (NC): LWA1 45, 50, 60, 70, 74, 80\, MHz, Landecker 150\,MHz, Haslam 408\,MHz and GMIMS 1383, 1418, 1456, 1487, 1499, 1521, 1614, 1625, 1660, 1700, 1712\,MHz data. All maps were smoothed to a common resolution of 5$^{\circ}$ and downgraded to N$_{{\rm{side}}}$ = 256. All instruments in this set are single-dish and so no angular scales over 5$^{\circ}$ are missing. \autoref{fig:radec} shows the empirical data employed for the coarse northern fit.       
    \item {\bf{Southern coarse}} (SC): Maipu/MU 45\,MHz, Landecker 150\,MHz, Haslam 408\,MHz, STAPS 1324, 1349, 1374, 1456, 1524, 1609, 1628, 1700, 1749, 1770\,MHz data. All maps were smoothed to a common resolution of 5$^{\circ}$ and downgraded to N$_{{\rm{side}}}$ = 256. All instruments in this set are single-dish and so no angular scales over 5$^{\circ}$ are missing. \autoref{fig:radec_south} shows the empirical data employed for the coarse southern fit.
    \item {\bf{Northern fine}} (NF): OVRO-LWA 41.8, 47, 52.2, 57.5, 62.7, 67.9, 73.2\,MHz, Haslam 408\,MHz and GMIMS 1383, 1418, 1456, 1487, 1499, 1521, 1614, 1625, 1660, 1700, 1712\,MHz data.  All maps were smoothed to a common resolution of 1$^{\circ}$ and downgraded to N$_{{\rm{side}}}$ = 256. \autoref{fig:radec_ovro} shows the OVRO-LWA data employed for the fine northern fit. This fit is only used to provide the 1 to 5$^{\circ}$ angular resolution information e.g the finer scaler anisotropies around the large scale patterns identified within the northern coarse fit.   
    \item {\bf{Southern fine}} (SF): Haslam 408\,MHz and STAPS 1324, 1349, 1374, 1456, 1524, 1609, 1628, 1700, 1749, 1770\,MHz data. All maps were smoothed to a common resolution of 1$^{\circ}$ and downgraded to N$_{{\rm{side}}}$ = 256. The lack of high resolution public data covering the southern hemisphere at frequencies lower than 408\,MHz results in a poorly constrained fit and so instead leaving the spectral index and amount of curvature to be completely free parameters, they were constrained at each pixel to be within $\pm5$ per cent of the value obtained in the Southern coarse fit. \autoref{fig:radec_south_fine} shows the empirical data employed for the fine southern fit. This fit is only used to provide the 1 to 5$^{\circ}$ angular resolution information e.g the finer scaler anisotropies around the large scale patterns identified within the southern coarse fit.   
\end{enumerate}

We find the least-squares per-pixel fit performs best (with the lowest reduced $\chi^{2}$) within log-space and with an additional free parameter ($f$) providing a multiplicative factor:

\begin{equation}
\label{eq:lsqfit}
\frac{T_{\rm{sync}}(p, \nu)}{T_{\rm{sync}}(p, \nu_{0})}  =  f \left(\frac{\nu}{\nu_{0}} \right )^{\beta_{\rm sync}},
\end{equation}
which, providing our model is adequate, should be very close to 1, or very close to 0 in log-space. This parameter is ancillary to help the fitting algorithm in dealing with pixels that contained systematic errors not being taken into account within our model, which is purely motivated by synchrotron emission behavior. \autoref{fig:fitex} shows an example of a per-pixel fit for the Northern coarse data set; the fitted parameters and reduced $\chi^{2}$ are given on the plot. The observational data across the available range of frequencies are simultaneously fit yielding a single spectral index and single curvature for each pixel. The per-pixel errors for each map are calculated as:
\begin{equation}
\sigma_{\rm{map}}(p) = \sqrt{{\sigma_{\rm{cal}}}(p)^2 + {\sigma_{\rm{offset}}}^2 + {\sigma_{\rm{thermal}}(p)}^2},
\end{equation}
where $\sigma_{\rm{cal}}(p)$ is the map calibration error percentage multiplied by the map temperature at pixel $p$, $\sigma_{\rm{offset}}$ is the single map offset error and $\sigma_{\rm{thermal}}(p)$ is a thermal noise error drawn from a random distribution centered around $0\,$K with a standard deviation equal to the map thermal noise levels given in \autoref{table:dat1}.

\subsection{Component separation }\label{sec:fgb}
In this section we present the complementary approach to fitting the synchrotron spectral parameters through a parametric component separation using \texttt{fgbuster} \citep{fgsoft, Stompor_2009, Stompor_2016}.  

 \fgb is  commonly adopted in recovering CMB data from multi-frequency observations $ d_{\nu}$ of multiple astrophysical components $s$, i.e.  the CMB together with the Galactic   foregrounds like thermal dust and synchrotron. Each signal scales  with frequency with   the so-called \emph{mixing matrix}, $A(\beta) $  depending on a set of spectral parameters, $\beta$: 
 \begin{displaymath}
     d_{\nu} = A({\beta})s +n_{\nu} 
 \end{displaymath}
 with $n_{\nu}$ being the instrumental noise of each frequency map. 
The methodology essentially consists in finding the optimal set of parameters $\beta$ and of component maps $s$, that  maximizes the spectral likelihood $\mathcal{L}$: 
\begin{equation} 
-2\ln\mathcal{L}(s,\beta)=(d-A(\beta)s)^T N^{-1} (d-A(\beta)s)+\mathrm{const.} ,\label{eq:like}
\end{equation}
with $N^{-1}$ being the inverse noise covariance matrix. 

 As the interstellar medium properties changes, the value of the Galactic spectral parameters  are expected to spatially  vary along multiple lines of sight.  This further complicates the recovery of the CMB polarized signals, as it might leave a mis-modeling bias into the faint \emph{B-}mode signal if those spatial variations are not properly taken into account.  However, latest results in the literature  \citep{Errard_2019, Puglisi_2022, rizzieri2025}, have shown that  the quality of the  recovered signals significantly improves both in terms of convergence   to the solution   and in the reduced level of mis-modelling bias when the spectral likelihood  is estimated  on a subset of pixels.  Practically, these sub-group of pixels  are optimally chosen  to be large enough to increase the SNR of each subset and small enough to reduce the mis-modelling  bias. 
 
 In this work, we employ \fgb for the first time to fit for both synchrotron and free-free emissions, the former modelled as a power-law with a non-zero running of the  spectral index (see \autoref{eq:ind} and \autoref{eq:curve}). We explicitly express the dependence of the emission by a given pixel $p$, $ \nu_0=45  $ MHz is the reference frequency (consistent with the parametric approach), whereas   $\nu_c$ is a parameter to be fitted and encodes the pivotal frequency for the $c_s$ parameter. We find that we need to generally set  $\nu_c\neq \nu_0\sim 70  $ GHz.   In fact, if  $\nu_c$ is constrained to a fixed value , e.g.$\nu_c =45 $ MHz, \fgb does not converge. This could be due to the fact that, although there is an extra-parameter to be estimated,  the \fgb likelihood model is better represented  with $\nu_c$ left to be free parameter,  giving more lever arm to the algorithm in estimating the synchrotron parameters.
Free-free emission  is modelled as :
\begin{equation}
s_{\rm{free}}(p, \nu) \propto  \left( \frac{\nu}{\nu_{0}} \right )^{\beta_f (p) }  ,
\end{equation}
 where $\beta_f$ is the spectral index, and  $\nu_0 = 408 $ MHz in this case. 
 
 Therefore, the set of  parameters fit by \fgb are thus : $\beta =\left[  \beta_s, c_s, \nu_c, \beta_f \right]$ and are assumed to be spatially variable in the sky. 

 Finally, as the frequency maps are pixillated at the common N$_{{\rm{side}}} =256$ resolution,  we perform the parameter estimation on pixel regions defined from the larger pixel locations selected from the N$_{{\rm{side}}} =64$ HEALPix grid. Thus, each  parameter estimate is obtained by considering a subset of 16 pixels.

 We run separately \fgb for  the 4  data sets  described in \autoref{sec:ff}, i.e.  encoding the North and South coarse and fine resolution maps. Each run was performed on a local machine and  took  50 minutes for the North maps, and 30 min for the South ones, (as the former data set presents a larger footprint than the latter). 

\subsection{Combination and point source removal}

The four different parameter fits result in four different maps for both the synchrotron spectral index and the spectral index curvature at the coarse and fine angular resolutions. It should be noted that as the value for $\beta_{s}$ changes across frequency in the model used in this work, we present maps of $\beta_{s}$ that are relative to the chosen values of $\nu_{0}$: 45\,MHz.

The large-scale fits were performed at $5^{\circ}$; the OVRO-LWA data were not included in this fit as they lack angular information on the sky larger than several degrees. The finer angular information in the northern hemisphere was provided by the OVRO-LWA data. The community are currently lacking the equivalent, high resolution, low frequency data covering the southern hemisphere. As the northern fit contains interferometric data, missing large angular scales and the southern fit only covers the frequency range 408 to 1170\,MHz the 1$^{\circ}$ parametric fit is only used to constrain the spectral index and curvature anisotropies. The coarse northern and southern fit were used to provide all the angular information above 5$^{\circ}$. We then joined together each north and south parameter maps. As we had more frequency data available in the  Northern surveys, we prioritized the northern results over the southern ones when considering the overlap region between the two. Regions which were not observed at all within the northern fit were supplemented with corresponding data from the southern maps.

\begin{figure}
    \centering
    \includegraphics[width=1\columnwidth]{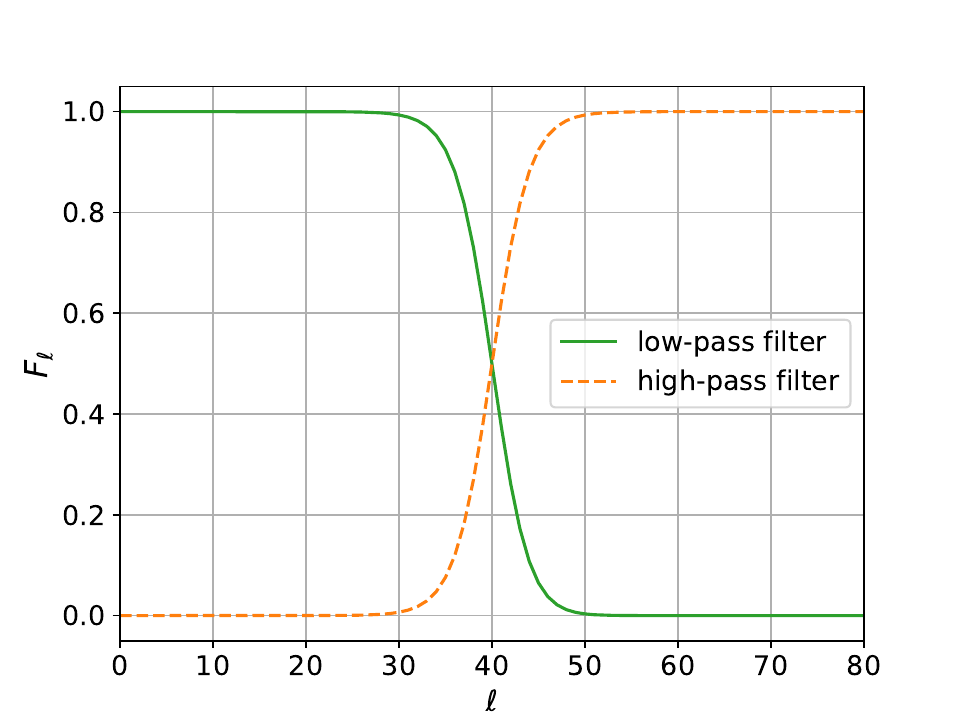}
    \caption{  (solid green)  the low-pass filter $F_{\ell}$ defined  in   \autoref{eq:filt}, (dashed orange)  the high-pass filter employed for the high resolution maps NF and SF, defined as $1-F_{\ell}$.  We particularly zoom-in on the region around $\ell_0$ to more clearly highlight the filter transition.}
    \label{fig:filter}
\end{figure}

\begin{figure*}
    \centering
    \includegraphics[width=1.75\columnwidth]{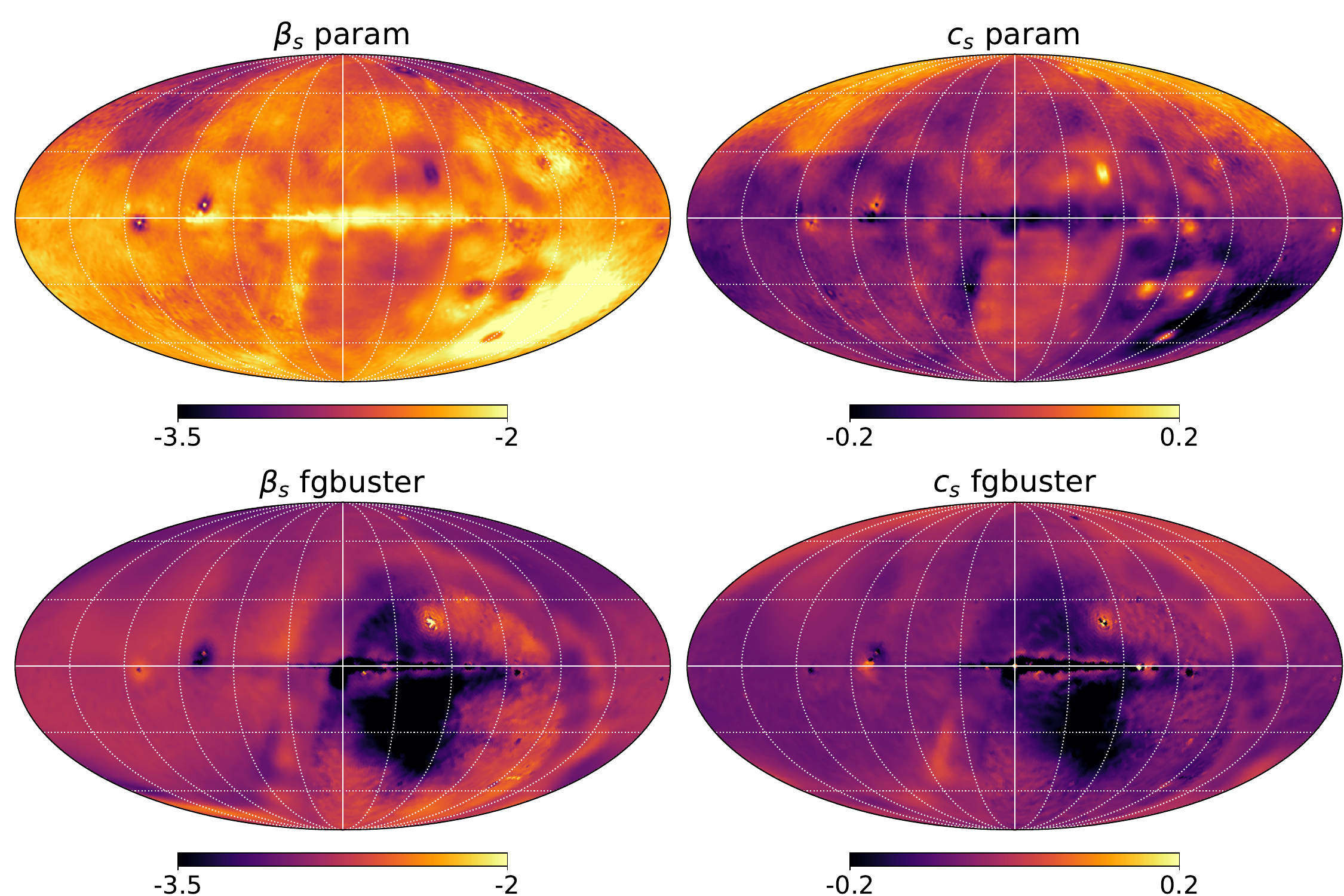}
    \caption{ Combined maps of NC,NF, SC and SF synchrotron spectral parameters (left panel) $\beta_s$ and (right panel) $c_s$ obtained with (top row) a parametric fit and  (bottom row) with \fgb. Note that $\beta_s$ is calculated at 45\,MHz.}
    \label{fig:finbeta}
\end{figure*}
\begin{figure*}
    \centering
    \includegraphics[width=1.75\columnwidth]{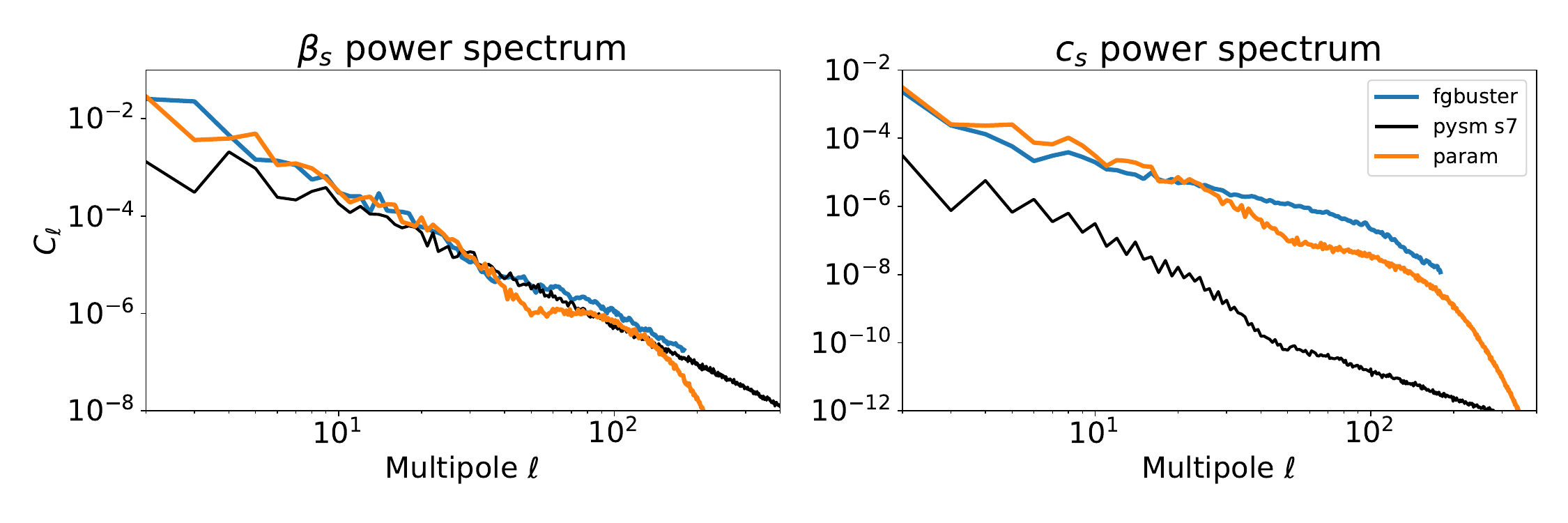}
    \caption{Angular power spectra estimated from  (left panel) $\beta_s$ and (right panel) $c_s$ full-sky maps shown in \autoref{fig:finbeta}. The spectra obtained  from the parametric fit and \fgb maps are respectively shown  in (solid orange) and (solid blue), in (solid black)  we also show   the power spectrum estimated from the template of the \texttt{pysm3} \texttt{s7} model.}
    \label{fig:spectra}
\end{figure*}

The joined maps lacked temperature information around the North and the South Celestial poles. Additionally, we wished to exclude extragalactic points source from our analysis. For the point source removal we made use of the Second {\emph{Planck}} catalog of compact sources \citep{planckPS} at 30\,GHz. Only sources with galactic latitudes ($b$) further than $7^{\circ}$ from $b = 0^{\circ}$ were excluded from the parameter maps to ensure that only extragalactic sources were removed. We filled these missing/masked locations with the \emph{diffusive inpainting} algorithm  presented in \citet{Puglisi_2020}.  The algorithm operates by filling in the pixels  with the mean value, which is calculated based on its closest neighboring pixels. This process is iteratively carried out, until the 2-norm difference of maps from two consecutive iterations reaches  an absolute tolerance of $10^{-4}$. For the maps shown in this work, less than 700 iterations are needed to achieve the set tolerance. 

Once both NC-SC and NF-SF maps were clipped together and inpainted we combined them into one single map employing the spherical harmonic decomposition routines available in the HEALPix library. We then low-(high-)pass filtered the coefficients from the coarse $a_{\ell m}^C $  (fine $a_{\ell m}^F $ ) maps as follows : 

\begin{equation} \label{eq:filter}
    a_{\ell m }^{tot}=  a_{\ell m }^{C} F_{\ell}  + a_{\ell m }^{F}  \left( 1 -F_{\ell}  \right) 
    ~~~, 
\end{equation}

with $F_{\ell}$ being a sigmoid filter:

\begin{equation}
    F_{\ell}  =  \left[1+  e^{ (\ell - \ell_0 ) / w  )}\right]^{-1}  
~~~,\label{eq:filt}
\end{equation}
with  $\ell_0=40$ and $w=2$ governing respectively the scale and the width   of the filter as a function of multipoles. We select the values of $\ell_0$ such that, on the one hand, we reliably low-pass filter the angular scales $\lesssim 5^\circ$ from our maps. On the other hand, we choose $w$ \emph{ad hoc} as the value that minimizes ringing artifacts caused by the filtering around bright sources and near the edges of the observational footprint.  The two employed filters are shown in \autoref{fig:filter},  in particular on the region around $\ell_0$ to more clearly highlight the filter transition. 

The top row of \autoref{fig:finbeta} shows the combined northern and southern, $1^{\circ}$ resolution synchrotron spectral index and curvature maps with point sources and missing data inpainted for the least-squares parametric fit. The spectral index at 45\,MHz generally fits the expected range ($-3.2$ to $-2.2$) for synchrotron emission, with the exception of those few pixels which present indices flatter than $-2.1$. A spectral index between $-2.2$ to $-2.1$ is generally expected for free-free emission and any values flatter than that are likely to just be associated with a poor fit. Whilst the reduced $\chi^{2}$ maps for each of the four individual fits (northern coarse and fine, southern coarse and fine displayed in \autoref{fig:res5} and \autoref{fig:res1},) show a good fit over all the pixels it is clear from the spectral index and curvature maps that these two parameters are highly degenerate and display a strong anti-correlation. The red areas within the spectral index map display an unusually flat spectral index and these exact same areas are shown in blue in the curvature map as they display an unusually large degree of curvature. The full covariance matrices of correlated errors are more revealing when determining the accuracy of the fit than just the reduced $\chi^{2}$ maps alone. These matrices are calculated as part of the fitting and so will be available within the software release associated with this work. As a visual guide to the areas within the spectral index and curvature maps which have the highest errors associated with them, we include \autoref{fig:param_abserr} which shows the spectral index and curvature absolute fractional errors (assuming a Gaussian propagation of errors between the coarse and fine parametric fits). Our least-squares spectral index map displays nonphysically shallow indices across regions measured by telescopes located the southern hemisphere. We have previously noted the lack of low resolution southern hemisphere data and it can clearly be seen from the fractional error maps that this is a problem for the accuracy of our parameter estimates. Additionally, the flatter spectral index values across the Galactic plane indicate that all the free-free emission within the empirical data may not have been removed by our free-free template. This limitation is not seen within the \fgb spectral index map and so future improvements to this work will focus on optimizing the joint emission fit.

\begin{figure}
 \centering  {\includegraphics[width=0.84\linewidth]{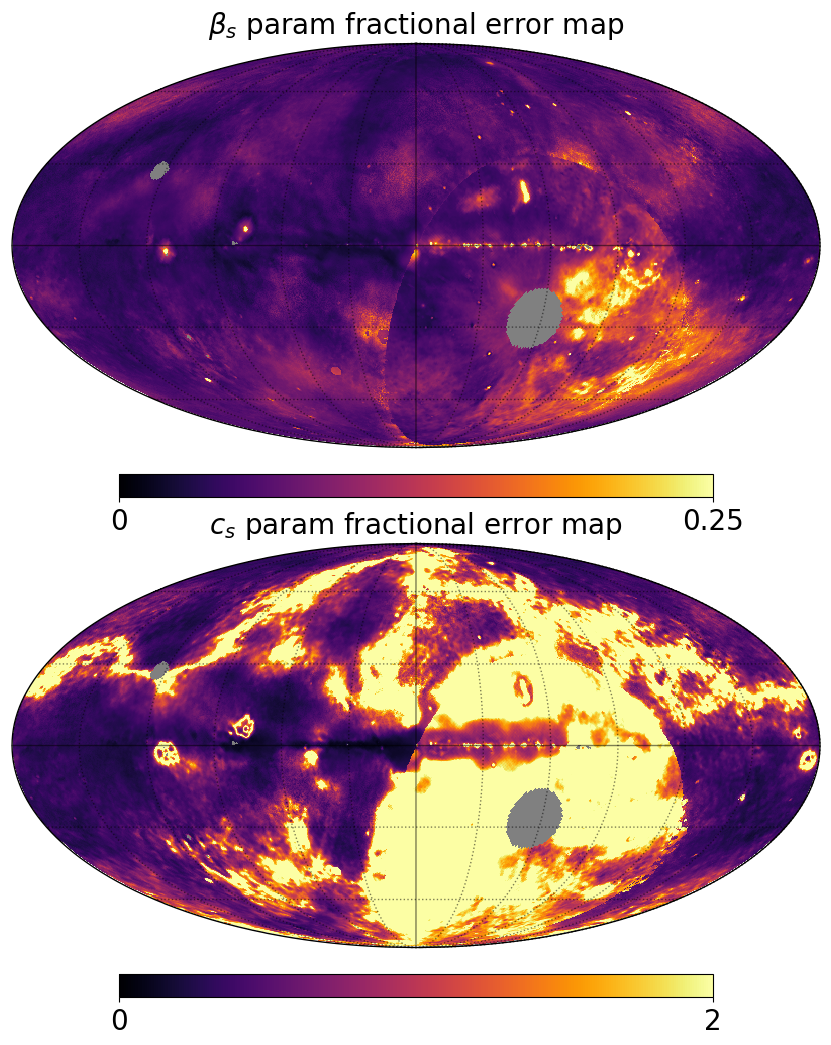}}\\
 \caption{Absolute fractional error maps for the least-squares parametric model of the synchrotron spectral index ({\emph{top}}) and the spectral index curvature ({\emph{bottom}}).}
 \label{fig:param_abserr}
 \end{figure}

\begin{figure*}
 \centering
{\includegraphics[width=0.185\linewidth]{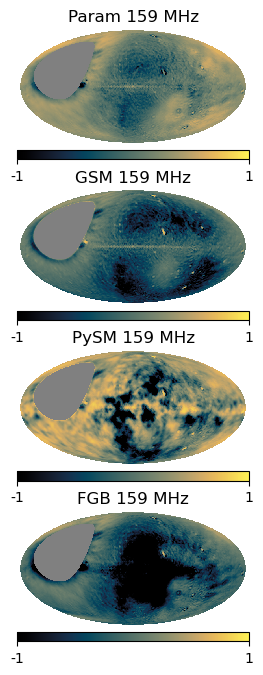}}
{\includegraphics[width=0.185\linewidth]{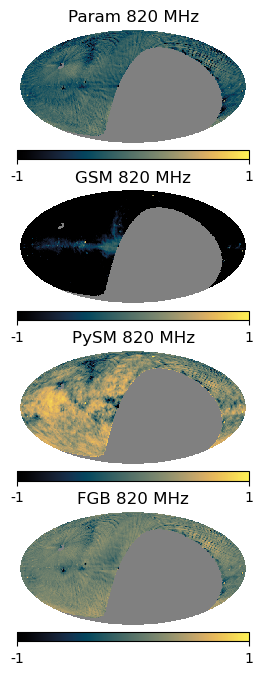}}
{\includegraphics[width=0.185\linewidth]{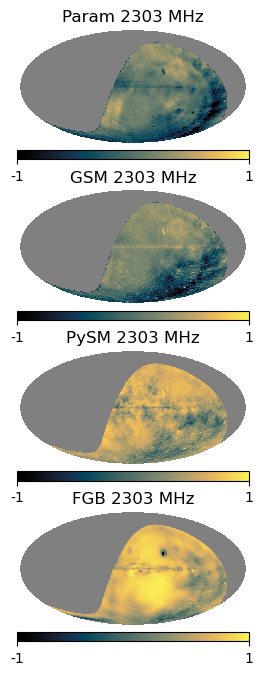}}
{\includegraphics[width=0.185\linewidth]{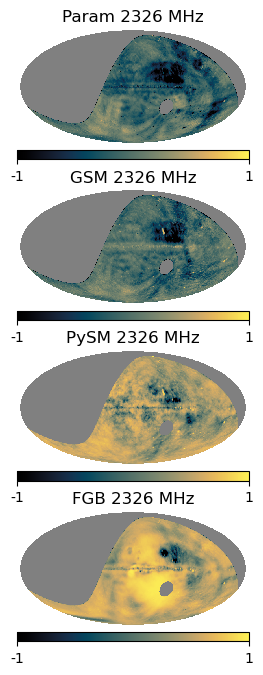}}
{\includegraphics[width=0.185\linewidth]{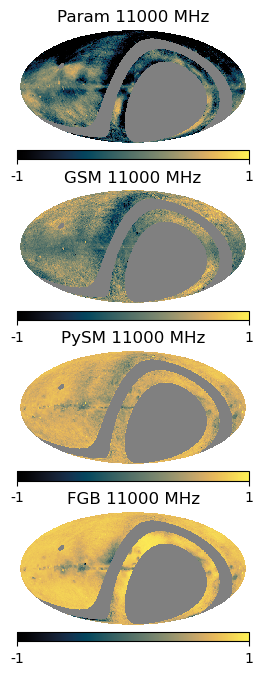}}
 \caption{Fractional difference maps between the empirical data and the models at 159, 820, 2303, 2326 and 11000\,MHz from left to right respectively. The top row shows the differences for the parametric model, the second row for the \texttt{GSM}, the third for the \texttt{pysm3} and the bottom row for \fgb .}
 \label{fig:diff_maps}
   \end{figure*}

In \autoref{fig:fgbres} we show the spectral parameters obtained with \fgb runs for the combined maps at both the low and high angular resolution. The final inpainted  maps are shown in  the bottom panel of \autoref{fig:finbeta}. 
We observe that the $\beta_s$ estimates are systematically steeper than those obtained from the parametric fit, particularly in regions near the  Galactic plane spreading to the southern hemisphere. A similar trend is evident in the $c_s$ map, where one can clearly distinguish the regions in which the northern and southern surveys are included. As indicated by the  correlation of the $\beta_s$ and $c_s$ maps and by the uncertainty maps (\autoref{fig:fgb_unc}), we notice that the correlation happens where the uncertainties are particularly large or unconstrained as the \fgb likelihood maximization does not converge. As expected,  this mainly  happens  in low SNR regions specifically at the southern  edges of the NC and NF surveys and in the region at $\ell, b \sim  (200^{\circ}, 45^{\circ}) $. The estimated spectral parameters are thus biased and unreliable in those regions.
   
Finally, we estimate the   power spectrum of angular correlations from   $\beta_s$ and   $c_s$  maps. We remark here that our maps employ large and small scales together. Moreover, the diffusive inpainting of both point sources and unobserved pixels (several percent of the sky) allowed us to perform a proper spherical harmonic decomposition and to mitigate the impact of extra-galactic sources. The angular power spectra\footnote{Estimated with Healpix routines \url{https://healpy.readthedocs.io/en/latest/healpy_spht.html}} obtained from the parametric least-squares fit and \fgb maps are shown in \autoref{fig:spectra} and are compared with the ones estimated from the template of the \texttt{pysm3} \texttt{s7} model, which uses the Haslam map as a synchrotron amplitude template, the model 4 synchrotron spectral index map from \citet{mamd} rescaled using S-PASS data and a spatially constant value for the spectral index curvature taken from \citet{kogut}. We notice  that the three $\beta_s$ power spectra follow almost  a similar power law scaling, although the spectrum estimated from the \texttt{pysm3} model is an order of magnitude lower at lower multipoles ($\ell \lesssim10$). The power spectra obtained with \fgb and the parametric fit instead clearly present the effects of having combined two maps with different angular resolutions, e.g. the loss of power at around $\ell \sim 40  $ and $\ell> 200$, corresponding respectively to the beam angular size of 5 and 1 $^{\circ}$. As the \fgb data products are pixellated at N$_{\rm side}=64$, the power spectra of \autoref{fig:spectra} in solid blue have been evaluated  at $\ell_{max} = 3N_{\rm side} -1$. The $c_s$ power spectra obtained both with \fgb and with the parametric fit show comparable amplitudes, both about 2 orders of magnitude larger than the \texttt{s7} model. Although the two maps have  almost similar full-sky average values:$-0.048,-0.073, -0.043$, respectively for the  parametric, \fgb and \texttt{s7} model, the discrepancy is due  to a larger variance of the estimated $c_s$ maps with respect to the \texttt{pysm3} one. We remark here that this  is somewhat  expected as the \texttt{s7} model has been derived from the Haslam synchrotron template matching the measured values reported in \citet{kogut} and has not been estimated from a fitting procedure.

\section{Results}
\label{sec:results}

The synchrotron spectral index and curvature parameter can be used alongside the free-free template employed in this work to construct an all-sky diffuse emission model below 2\,GHz. $T_{\rm{ff}}(p, \nu)$ is calculated as in \autoref{sec:ff} and $T_{\rm{sync}}(p, \nu)$ is calculated as in \autoref{eq:ind} and \autoref{eq:curve} and $\nu_{0}$ is chosen to be 408\,MHz with the synchrotron amplitude template at 408\,MHz provided by the 56 arcmin Haslam data. $\beta_\text{s}$ and $c_s$ are the fitted spectral index and curvature map at $1^{\circ}$ resolution. 

\begin{table}
\begin{center}
\addtolength{\tabcolsep}{-0.55em}
\begin{tabular}{c | c c c c c} 
 \hline
 & 159\,MHz & 820\,MHz  & 2303\,MHz & 2326 \,MHz & 11000\,MHz \\
 \hline\hline
 {\bf{Parametric}} & 14 - 39 & 9 - 29 & 13 - 39 &9 - 32 & 21 - 100 \\
  {\bf{\texttt{GSM}}} & 12 - 57 & 172 - 343 & 17 - 39 & 6 - 24 & 18 - 47 \\
   {\bf{\texttt{pysm3}}} & 22 - 58 & 18 - 50 & 34 - 67 & 36 - 60 & 57 - 74 \\
    {\bf{\fgb}} & 12 - 56  & 10 - 24  & 46 - 83 & 48 - 72 & 75 - 87  \\
 \hline
  \hline
\end{tabular}
\caption{Percentage difference ($25^{\rm{th}}$ and $75^{\rm{th}}$ percentile) between sky models and empirical data at five frequencies. The empirical maps cover different regions of the sky but the Galactic plane is always excluded from the difference calculation.}
 \label{tab:modcomp}
\end{center}
\end{table}
 
\begin{figure*}
 \centering
  {\includegraphics[width=0.81\linewidth]{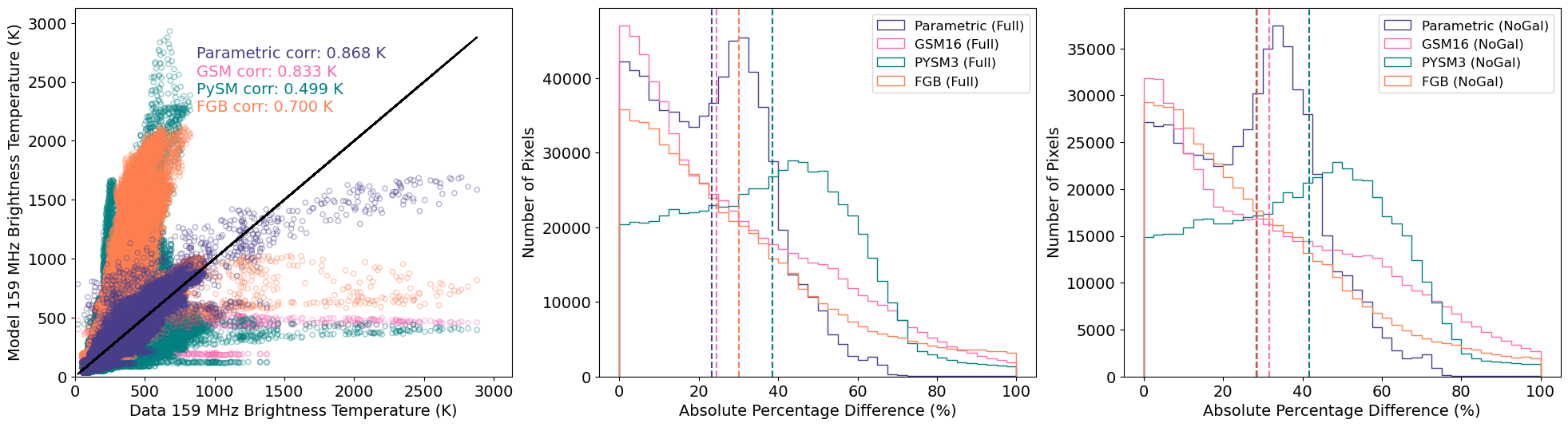}}\\
  {\includegraphics[width=0.81\linewidth]{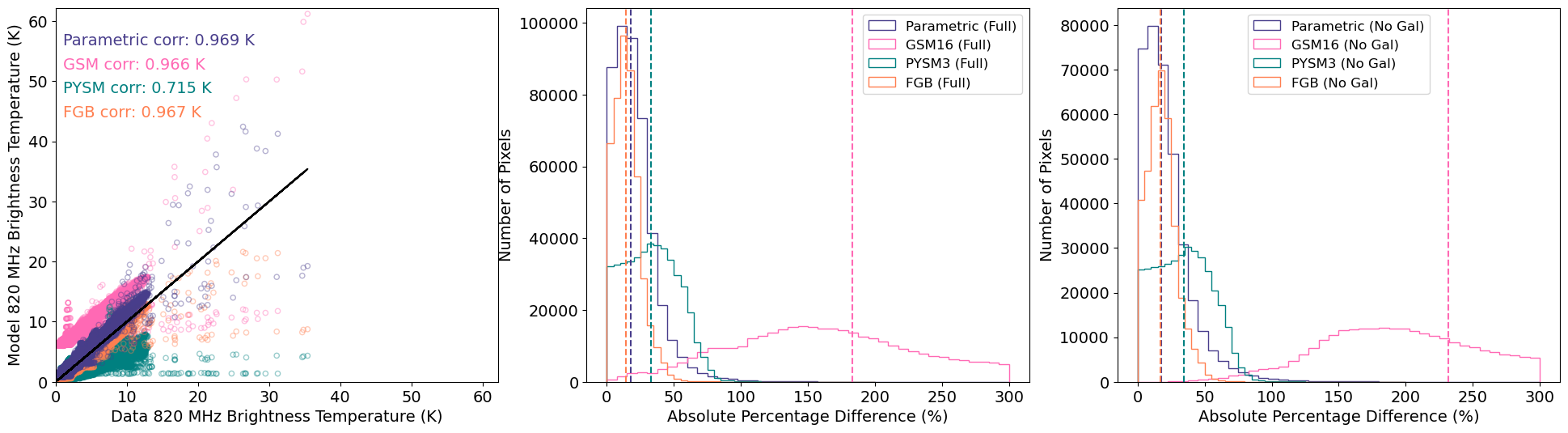}}\\
  {\includegraphics[width=0.81\linewidth]{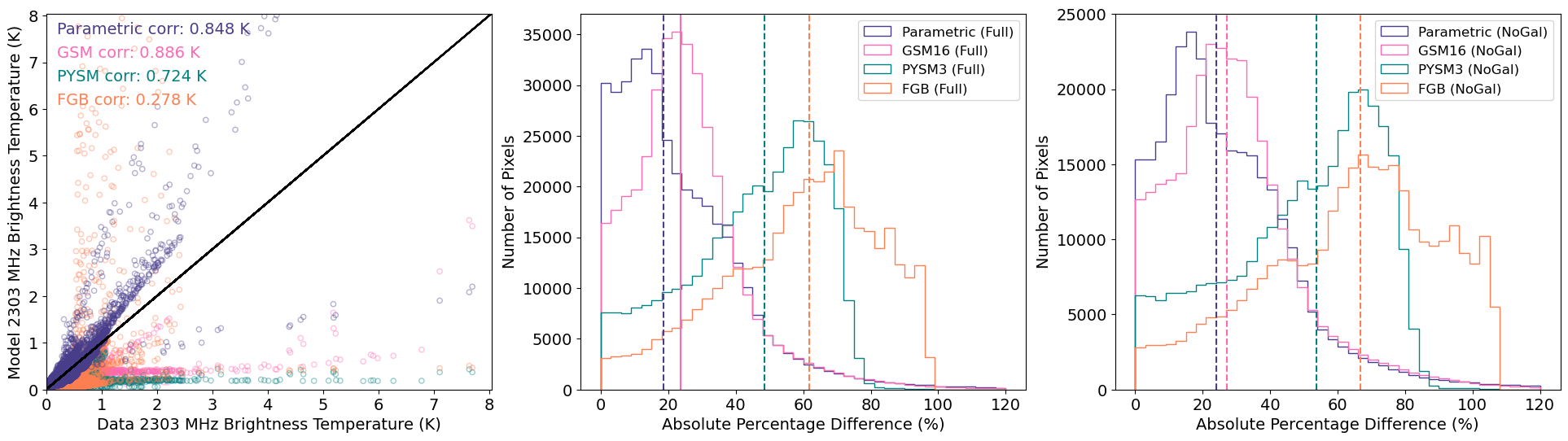}}\\
  {\includegraphics[width=0.81\linewidth]{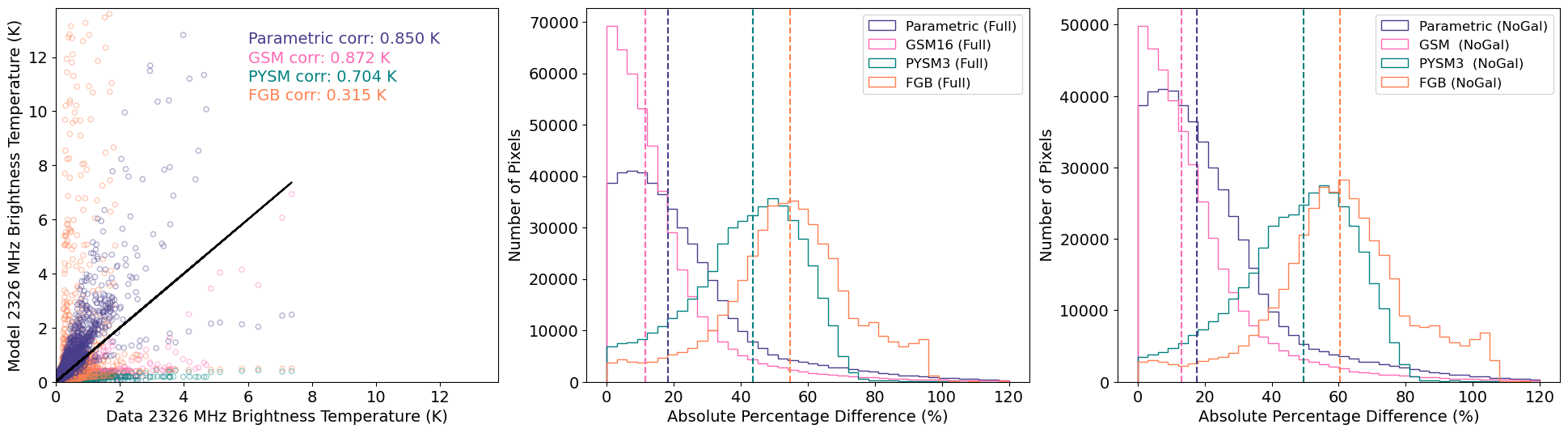}}\\
  {\includegraphics[width=0.81\linewidth]{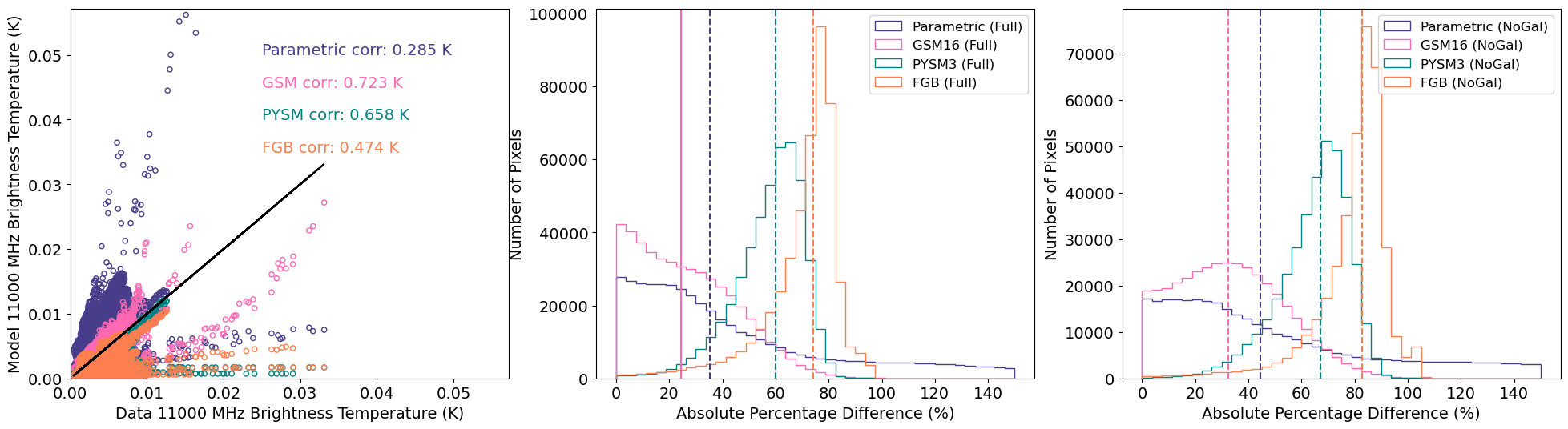}}\\
 \caption{First to last row: correlation and histogram comparison between various sky models and empirical partial sky data at 159, 820, 2303, 2326, 11000\,MHz, respectively. The scatter plots exclude the Galactic plane region.}
 \label{fig:comp}
   \end{figure*}

To validate our sky model we enlist the use of public partial sky data which were not used within our fit: EDA2, Dwingeloo, Rhodes, S-PASS and QUIJOTE. The EDA2 survey is a southern survey at 159\,MHz with a resolution of $3.1^{\circ}$ \citep{EDA2}, Dwingeloo surveyed the northern hemisphere at 820\,MHz with a resolution of $1.2^{\circ}$ \citep{dwing}, the Rhodes southern hemisphere survey \citep{jonas} observed at a resolution of 20 arcmin at 2326\,MHz, S-PASS conducted a 2303\,MHz survey with a resolution of 8.9 arcmin \citep{spass} and QUIJOTE MFI are observing the northern sky at 11, 13, 17 and 19\,GHz with a resolution of $1.0^{\circ}$ \citep{q11}.  

We compare our diffuse sky model estimates with those of the \texttt{GSM} and \texttt{pysm3}. For the \texttt{GSM} we use the high resolution model which produces sky estimates at 48 arcmin; as the \texttt{GSM} represents total diffuse emission we do not need to include anything else. This is the same for the \texttt{fgbuster} sky model estimates which already include both synchrotron and free-free emission. For the \texttt{pysm3} we use the `\texttt{s7}' synchrotron model. Both the \texttt{pysm3} and the least-squares parametric fit give synchrotron emission models so we add to them the free-free emission template (scaled to the appropriate frequency) from \autoref{sec:ff}. The model/data comparisons are performed at resolutions of $3.1^{\circ}$ for the EDA2 data, $1.2^{\circ}$ for the Dwingeloo data and at $1.0^{\circ}$ for the Jonas, S-PASS and QUIJOTE data.

\begin{figure}
    \centering
    \includegraphics[width=0.9\columnwidth]{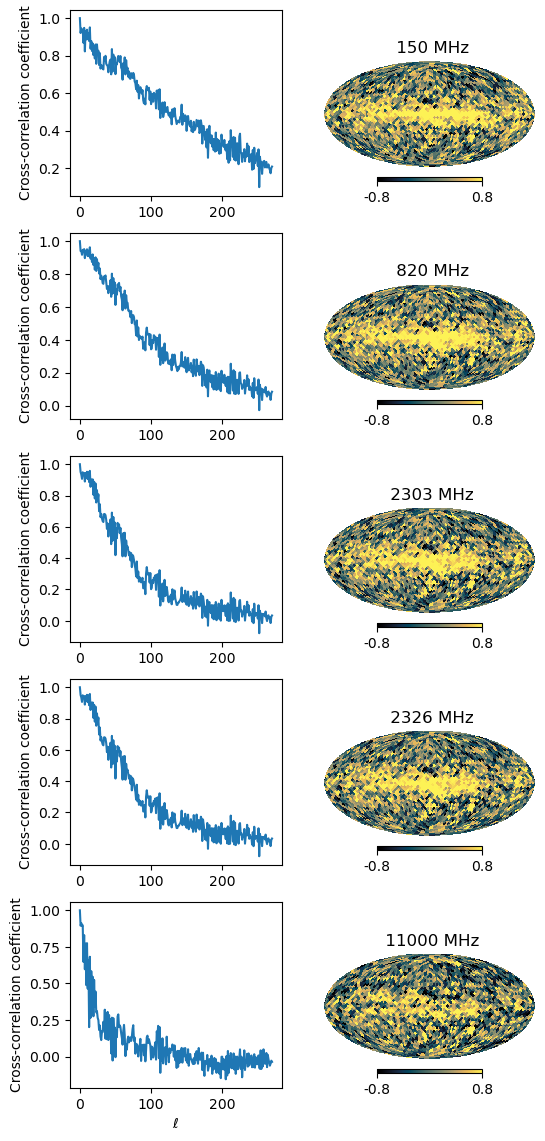}
    \caption{Correlation coefficient plots and maps between the parametric synchrotron estimate and free-free template at 159, 820, 2303, 2326 and 11000\,MHz. All maps were made at 1$^{\circ}$ resolution. The plots show correlation across angular harmonics whilst the maps show the correlation coefficient within super-pixels.}
    \label{fig:ff_corr}
\end{figure}

\begin{figure}
    \centering
    \includegraphics[width=0.75\columnwidth]{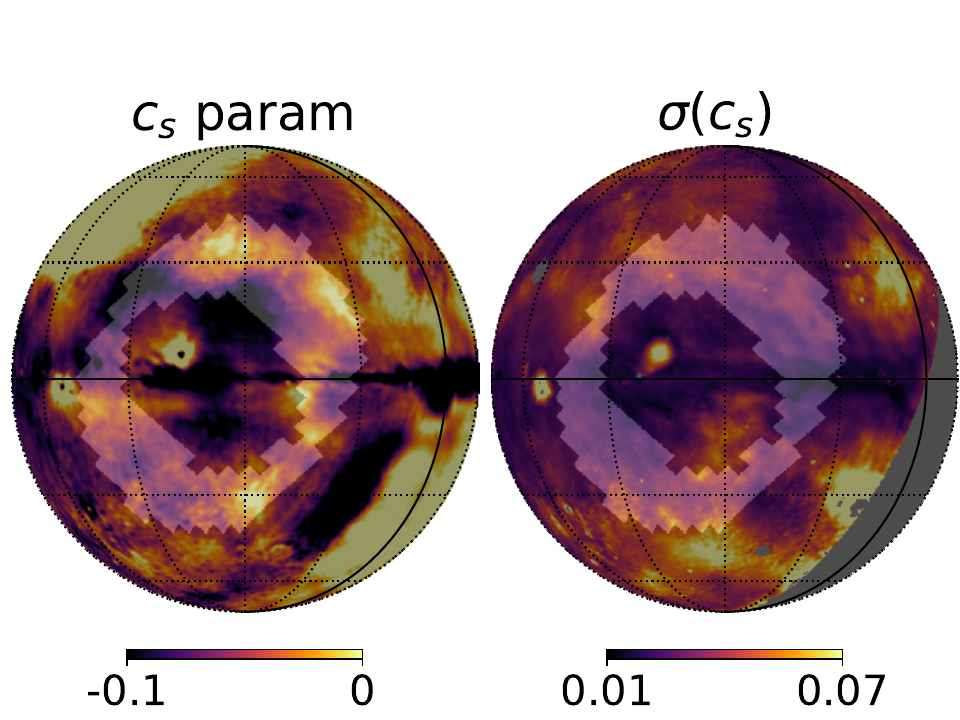}
    \caption{$c_s$ (left) and $\sigma(c_s)$ maps obtained with the parametric least-squared fit shown within the ARCADE2 observational footprint \citep{kogut} and in Orthografic projection centered in $\ell, b= (60^{\circ},0^{\circ} )$.  For the sake of comparison,  we set the color-scale of the $c_s$ map to be the same as the one in \citep[Fig.4]{kogut}.}
    \label{fig:cs_arc}
\end{figure}

\autoref{fig:diff_maps} shows the fractional difference maps between the diffuse sky models and the data defined as:
\begin{displaymath}
    \delta = \frac{m_{\rm data} - m_{\rm model}}{m_{\rm data}}, 
\end{displaymath}

with $m_{\rm data}$ and $m_{\rm model}$ being respectively the map observed and the one obtained by evaluating \autoref{eq:ind} from the fit spectral parameters at the  same frequency as the data.
 In order for the empirical data to only contain diffuse and compact emissions the constant map zero-level was calculated and removed using the technique described in \autoref{sec:zero}. The difference maps are arranged by frequency from left to right (159, 820, 2303, 2326 and 11000\,MHz) and by model from top to bottom (Parametric, \texttt{GSM}, \texttt{pysm3} and \texttt{fgbuster}). Interestingly each model struggles to replicate the empirical data across different regions of the sky as frequency changes; there is no consistent problem area, e.g. as in the Galactic plane. The \texttt{GSM} provides a good sky model except at 820\,MHz. In particular, we anticipate that the \texttt{s7} model implemented in \texttt{pysm3} would have exhibited larger residuals when compared against data at frequencies $<1\,\mathrm{GHz}$. This is because \citet{pysm3} specifically validated the reliability of this model only for frequencies above this threshold, implying that its performance may degrade when extrapolated to lower-frequency regimes. This could explain the reason of larger residuals at 159 GHz.} The parametric model performs well at all frequencies except at 11000\,MHz at high galactic latitudes; this motivates our decision to recommend the parametric sky model for estimating the sky between 45 and 2300\,MHz. The \fgb model can be seen to overestimate the sky at low frequencies and underestimate it at higher frequencies, though it models the data at 820\,MHz very well.        

In \autoref{fig:comp} we further quantify the differences between the empirical data and sky models through the use of correlation plots and histogram distributions. As sky models are most typically used outside of the Galactic plane within the fields of CMB and 21\,cm cosmology we choose to mask out the Galactic plane from some of our results. This is done through the use of the \emph{Planck} GAL080 mask \footnote{\url{https://pla.esac.esa.int}} which masks out the Galactic plane, leaving 80 per cent of the sky pixels unmasked. The left-hand column shows correlation plots between the data and models outside of the Galactic plane mask. The correlation plots feature a black line of gradient 1 to visually highlight the slope a perfect model would have. The parametric model has the highest correlation coefficient with the data at 159 and 820\,MHz whilst the \texttt{GSM} has the highest correlation coefficient at 2303, 2326 and 11000\,MHz. The middle and right-hand columns show histogram distributions of the absolute percentage differences between the model and data for all four models, in the middle column the full available sky is used whereas the Galactic plane is masked out for the right-hand column plots. The median absolute percentage differences are marked on the histograms using dotted lines. \autoref{tab:modcomp} is included to aid the interpretation of the histograms in \autoref{fig:comp} as it shows the 25$^{\rm{th}}$ and 75$^{\rm{th}}$ percentiles between the data and the four models when the Galactic plane has been excluded from the calculation. At 159 and 2303\,MHz the parametric model provides lowest median difference between the data and the empirical model, at 820\,MHz the \fgb model has the lowest median difference and at 2326 and 11000\,MHz the \texttt{GSM} provides the lowest median difference between the data and the empirical mode.

In \autoref{fig:ff_corr} we assess the level of correlation between our synchrotron estimates at each frequency and the free-free emission template used in this work. We expect a strong correlation across the largest angular scales as all diffuse Galactic emissions show the Galactic plane as the brightest region in the Galaxy. However, we then expect the correlation to drop off across medium to small angular scales. The plots show the correlation coefficient across angular harmonic scales: 
\begin{equation}
\rho_{\ell} = \frac{C_{\ell} ^{\rm{sy, \, ff}}}{\sqrt{C_{\ell} ^{{\rm{sy, \, sy}}} C_{\ell} ^{{\rm{ff, \, ff}}}}},
\end{equation}

where $C_{\ell} ^{{\rm{sy, \, sy}}}$ is the synchrotron model auto-correlation power spectrum and $C_{\ell} ^{{\rm{sy, \, ff}}}$  is the cross-correlation power between the synchrotron and free-free models. For the maps we split the sky into sub-regions of 1024 pixels (N$_{\rm{side}}$ = 8 out of the full sky which is at N$_{\rm{side}} = 256$.) and plot the Pearson correlation coefficient between the synchrotron and the free-free model for the pixels within the region. From the maps we can see that the strongest correlation between our synchrotron and free-free emission templates occurs across the Galactic plane, as expected. As the frequency increases the synchrotron emission across the full sky drops and so the correlation with free-free emission across the Galactic plane becomes weaker.

We finally compare the curvature parameter with the one estimated by \citet{kogut}. They combined radio surveys at 22\,MHz,  1.4, 3 and 10\,GHz, and observed a steepening in the radio spectrum  corresponding to a curvature parameter of $c_s = -0.052\pm 0.005$ as estimated within the ARCADE2 footprint.  \autoref{fig:cs_arc} shows the estimated $c_s$ obtained from the parametric fit. Therefore, within the observational patch defined by \citet{kogut}, we compute the mean value of $c_s$ using weights given by the inverse of the total uncertainty, $\sigma(c_s)$, where $\sigma(c_s)$ is obtained as the quadratic sum of the errors from the coarse and fine estimates (see \autoref{fig:res5} and \autoref{fig:res1}). We obtain $c_s = -0.0517 \pm 0.0007$,   compatible, albeit more accurate, with the value reported by \citet{kogut}.

\section{Conclusions}
\label{sec:conc}
We present the first publicly available full-sky maps of the diffuse synchrotron spectral index $\beta_s$ at 45\,MHz and spectral index curvature $c_s$ , both at N$_{{\rm{side}}}= 256$ at the angular  resolution of 1$^{\circ}$. To the best of our knowledge, these estimates are unprecedented: prior works \cite{mamd, Krachmalnicoff_2018} relied on $\beta_s$ at a coarser 5$^{\circ}$ resolution  and covered only a more restricted frequency range.

These results were obtained through a parametric fit of 37 empirical sky maps. These maps differed in sky coverage, namely northern and southern hemisphere, as well as angular resolution. Only two of the maps used were made using instrumentation capable of measuring the absolute zero-level. This work was made possible thanks to the wealth of high quality radio data that has been made public, enormous progress in the production of free-free emission templates and the analytical power of a simple linear regression technique: the temperature-temperature plot. Four separate fits were performed: the northern 5$^{\circ}$, the southern 5$^{\circ}$, the northern 1$^{\circ}$ and the southern 1$^{\circ}$ fits. The northern and southern hemispheres were joined together, favoring the northern data in the overlap region, and extragalactic point sources were masked. Masked pixels and missing pixels due to scan strategies were inpainted and finally the coarse and fine angular scales were combined in spherical harmonic space.

Within the frequency range of 45 to 2300\,MHz free-free emission is not completely negligible across the full sky and so it must be accounted for in our analysis. We tested two approaches for handling the free-free emission. The first approach removes a free-free template at each frequency and performs a least-squares regression between the synchrotron model and the empirical data at each pixel. The second approach separates  the synchrotron and free-free   components with a Bayesian parametric algorithm \texttt{fgbuster}, commonly adopted by the community for  CMB foreground cleaning.

The spectral index and curvature maps were used to predict diffuse Galactic synchrotron emission at 159, 820, 2303 and 2326 and 11000\,MHz using the Haslam 408\,MHz all-sky map as a synchrotron amplitude template. We \emph{ad hoc} chose these five frequencies as they span a wide range of interest for both the 21\,cm and CMB communities.  Moreover, these data have not been included in our fit, due to their partial sky coverage. We therefore leverage this to further test and validate the models obtained with the two approaches. 

Through this empirical evaluation, it is evident that the combination of free–free template subtraction and a least-squares fitting procedure yielded better results than the alternative approach based on component separation, the latter being more strongly affected by instrumental systematics arising from calibration errors, which in turn led to an underestimation of the associated uncertainties and larger $\chi^2$.

We plan to devote future work to overcoming this limitation,  by  fitting spectral parameters together with calibration  errors and offsets within the same framework, employing  novel methodologies recently described in the literature \citep{bayes, beyondplanck}. A joint fitting approach will enable a proper assessment of the anti-correlation we see between the least-squares parametric spectral index and curvature values, as this may be physical or may simply be due to degeneracies within our fit. 

We compare our emission models with those of the \texttt{pysm3} and \texttt{GSM}. At the test frequencies, we evaluate that  the least-squares parametric fit provides the closest model to the data at 159, 820 and 2303\,MHz, coming second to the \texttt{GSM} at 2326\,MHz. Our least-squares parametric model for synchrotron emission provides the most reliable sky model between 45 and 2300\,MHz as it consistently predicts the sky temperature to accuracies of around 20 percent on average whilst over methods jump between 10 and 70 per cent in terms of average accuracies. We, therefore, recommend the community to use the templates obtained with  the parametric  model as it shows higher level of accuracy. By 11\,GHz the model starts to show significant deviations from the empirical QUIJOTE data and so this model is not yet complete. However, the power of the parametric framework presented in this paper is that as new data within the 50 to 10000\,MHz (synchrotron and free-free dominated) frequency regime become publicly available (e.g. C-BASS \citep{mike}) this sky model can and will be updated and improved.   

\section*{Acknowledgments}

 We thank: Brandon Hensley, Susan Clark, Paddy Leahy, Mathieu Remazeilles, Shamik Ghosh,  Jacques Delabrouille, Roke Cepeda-Arroita and the Pan-Experiment Galactic Science group for the comments and fruitful discussion and suggestions  on synchrotron modelling. GP thanks Arianna Rizzieri and Josquin Errard for useful insights and suggestions in parametric component separation.
 We also thank the excellent resource of the LAMBDA Legacy Archive \footnote{\url{https://lambda.gsfc.nasa.gov/}} which houses the majority of the sky maps used in this work. 
 
GP acknowledges financial support under the National Recovery and Resilience Plan (NRRP), Mission 4, Component 2, Investment 1.1, Call for tender No. 104 published on 2.2.2022 by the Italian Ministry of University and Research (MUR), funded by the European Union – NextGenerationEU– Project Title ``SHIFT'' – CUP 55723062008 - Grant Assignment Decree No. 962 adopted on June 30th 2023 by the Italian Ministry of Ministry of University and Research (MUR).

Some of the results in this paper have been derived using the following \texttt{healpy} and \texttt{HEALPix} packages.

\section*{Data Availability}

We publicly release  the  code and the scripts used  produce the results presented in this paper at \url{https://github.com/giuspugl/fitting_synchrotron}.   At the   same repository, it is possible to download  the maps of $\beta_s$ and $c_s$ described in this work.



\bibliographystyle{mnras}
\bibliography{refs} 




\appendix

\section{Empirical data}

\begin{table}
\begin{center}
\addtolength{\tabcolsep}{-0.55em}
\begin{tabular}{||c c c||} 
 \hline
Survey & Freq. (MHz) & Offset (K) \\ [0.5ex] 
 \hline\hline
 Maipu/MU  & 45 & $3063\pm112$\\
  \hline
LWA1 & 45 &  $2861\pm94$ \\
  \hline
  LWA1 & 50 &  $2290\pm70$ \\
   \hline
   LWA1 & 60 & $1608\pm42$\\ 
 \hline
 LWA1 & 70 & $1127\pm26$ \\
 \hline
 LWA1 & 74 & $1026\pm22$ \\
 \hline
 LWA1 & 80 &   $845\pm19$ \\
  \hline
Landecker  & 150 &  $84\pm6$\\
 \hline
GMIMS-HBN & 1383 & $-0.39 \pm 0.02$ \\
 \hline
GMIMS-HBN & 1418 & $-0.32\pm0.01$ \\
\hline
GMIMS-HBN & 1456 &   $-0.31\pm0.01$\\
\hline
GMIMS-HBN & 1487 & $-0.31\pm0.1$\\
\hline
GMIMS-HBN & 1499 &   $-0.31\pm0.1$\\
\hline
GMIMS-HBN & 1521 & $-0.33\pm0.01$ \\
\hline
GMIMS-HBN & 1614 & $-0.27\pm0.01$  \\
\hline
GMIMS-HBN & 1625 & $-0.24\pm0.01$ \\
\hline
GMIMS-HBN & 1660 & $-0.26\pm0.01$ \\
\hline
GMIMS-HBN & 1700 &  $-0.22\pm0.01$\\
\hline
GMIMS-HBN & 1712 & $-0.022\pm0.01$ \\
\hline
GMIMS-STAPS & 1324 & $-0.83\pm0.01$ \\
\hline
GMIMS-STAPS & 1349 & , $-0.77\pm0.01$ \\
\hline
GMIMS-STAPS & 1374 &  $-0.73\pm0.01$\\
\hline
GMIMS-STAPS & 1456 & $-0.63\pm0.01$\\
\hline
GMIMS-STAPS & 1524 & $-0.57\pm0.01$ \\
\hline
GMIMS-STAPS & 1609 & $-0.49\pm0.01$ \\
\hline
GMIMS-STAPS & 1628 &  $-0.49\pm0.01$\\
\hline
GMIMS-STAPS & 1700 & $-0.45\pm0.01$\\
\hline
GMIMS-STAPS & 1749 &  $-0.43\pm0.01$ \\
\hline
GMIMS-STAPS & 1770 &  $-0.41\pm0.01$\\
\hline
  \hline
\end{tabular}
\caption{The calculated offset values for each of the single-dish empirical data sets used in this work.}
 \label{tab:offvals}
\end{center}
\end{table}

      \begin{figure*}
 \centering
{\includegraphics[width=0.99\linewidth]{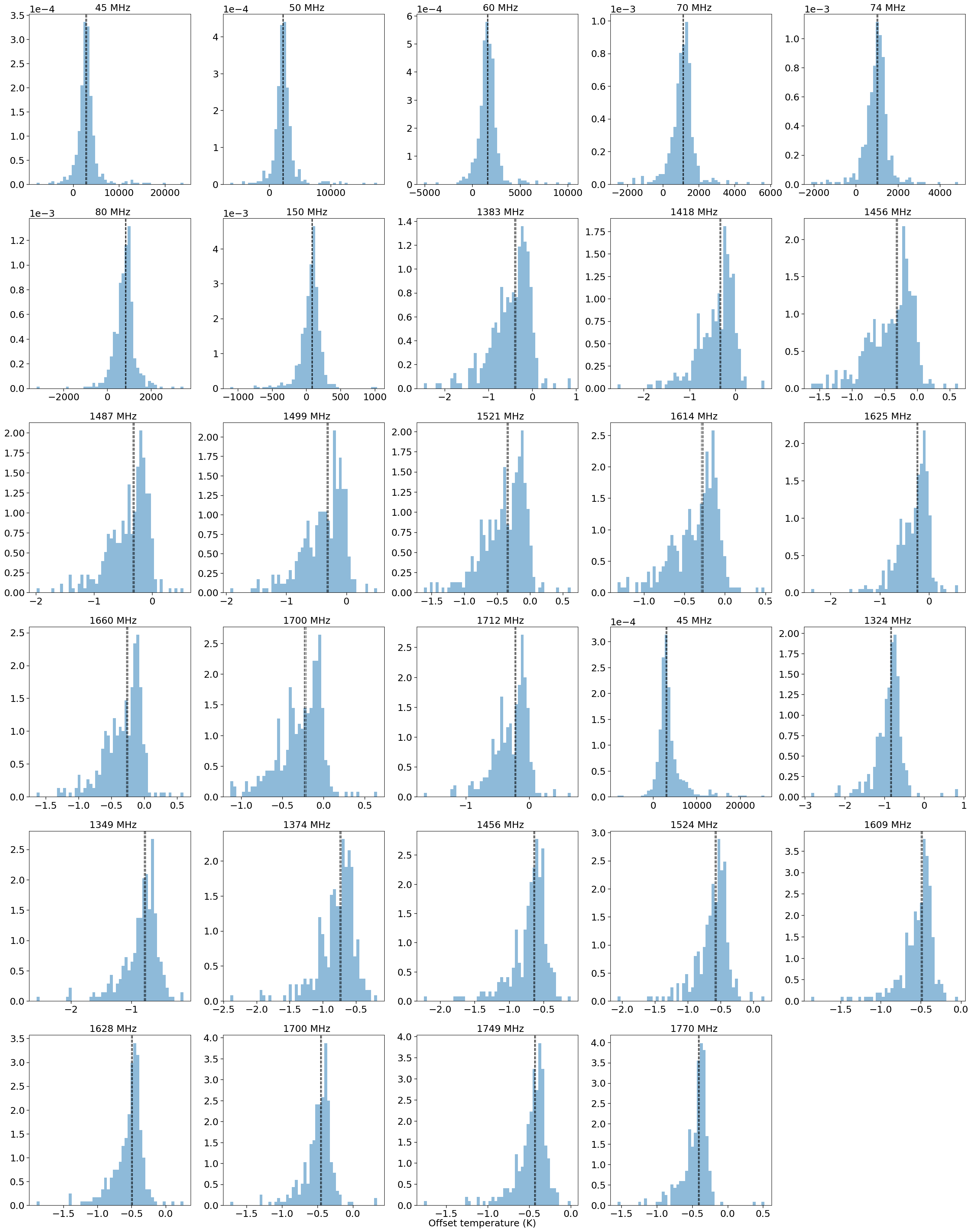}}\\
 \caption{Histogram distributions of the offset values in Kelvin; the mean (solid black line) and standard error on the mean (dashed black line) are used as the map offset value and uncertainty.}
 \label{fig:ohists}
   \end{figure*}
   \begin{figure*}
 \centering
{\includegraphics[width=0.95\linewidth]{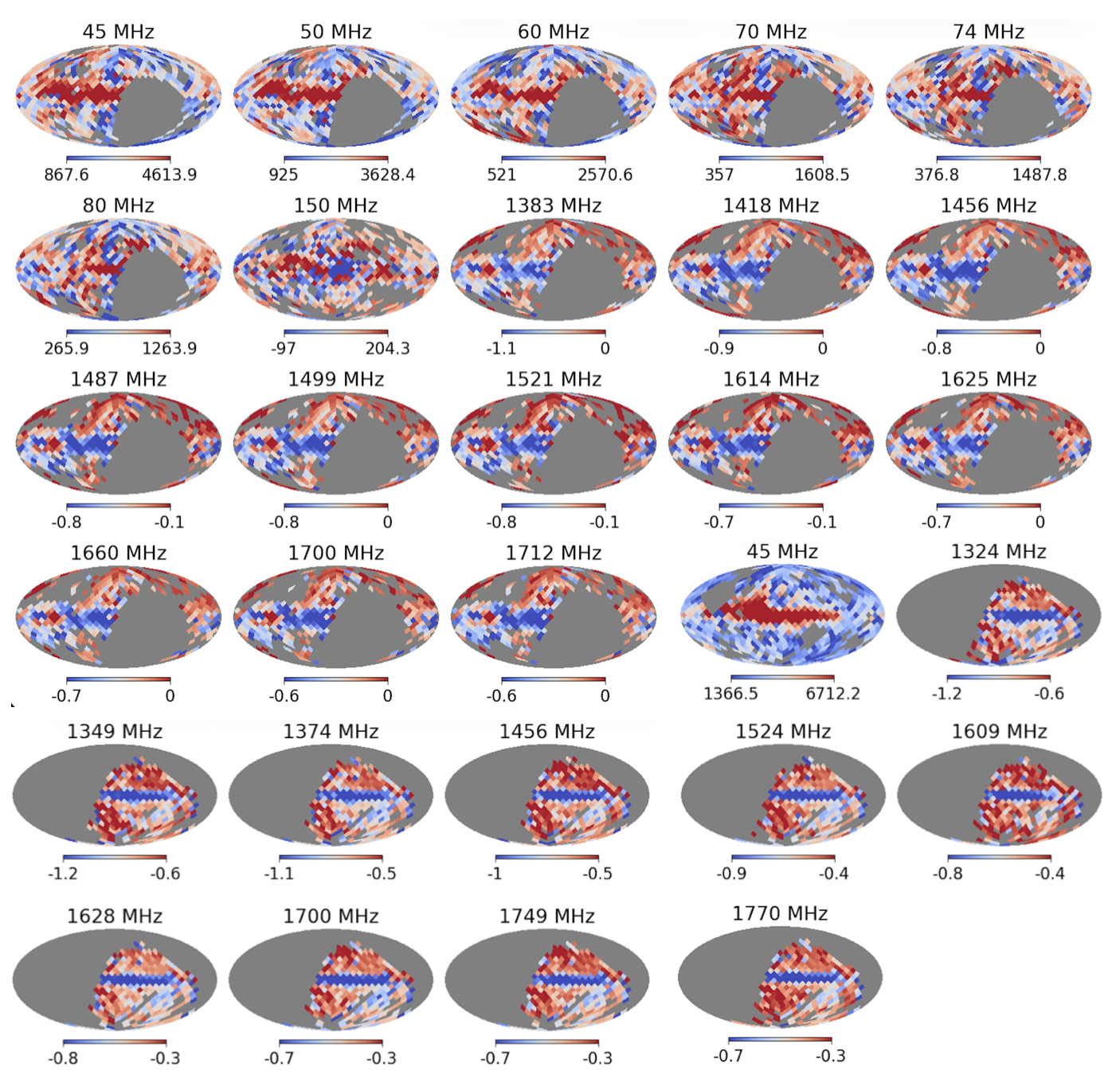}}\\
 \caption{Maps of the offset values calculated within each super pixel for each frequency map. Color scale is in K.}
 \label{fig:omaps}
   \end{figure*}

We report in \autoref{tab:offvals} the offset values that have been removed to each maps employed  before the parameter fitting. We show in \autoref{fig:ohists} the distribution of the estimated offsets together with the maps of the offsets (\autoref{fig:omaps}) estimated in a lower pixel grid  $N_{\rm side}=8$ . 

\begin{figure*}
 \centering
  {\includegraphics[width=0.85\linewidth]{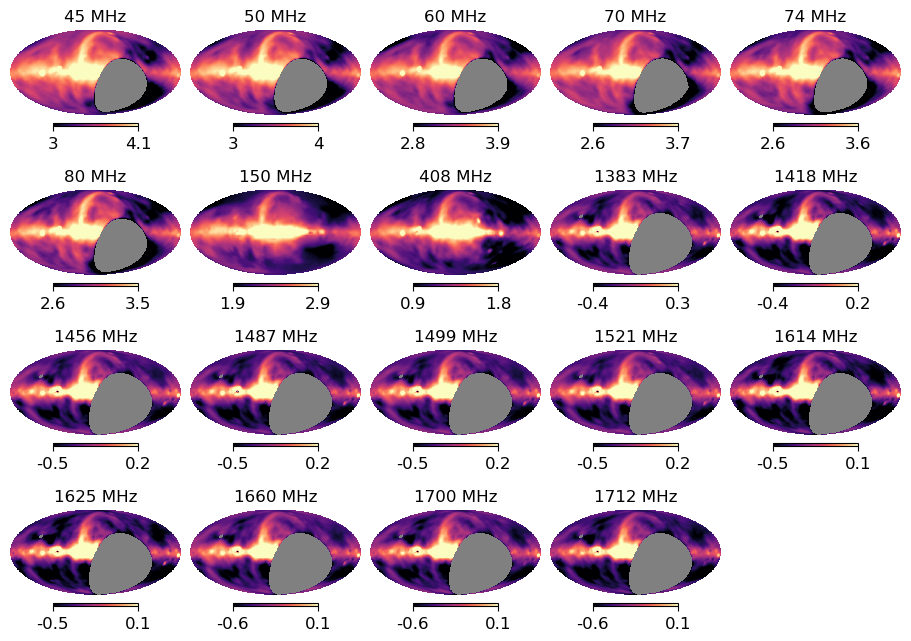}}\\
 \caption{Frequency maps at 5$^{\circ}$  resolution employed in the fit of $\beta_s$ and $c_s$ across the Northern hemisphere. Color scale is log$_{10}$(K).}
 \label{fig:radec}
   \end{figure*}

\begin{figure*}
 \centering
{\includegraphics[width=0.85\linewidth]{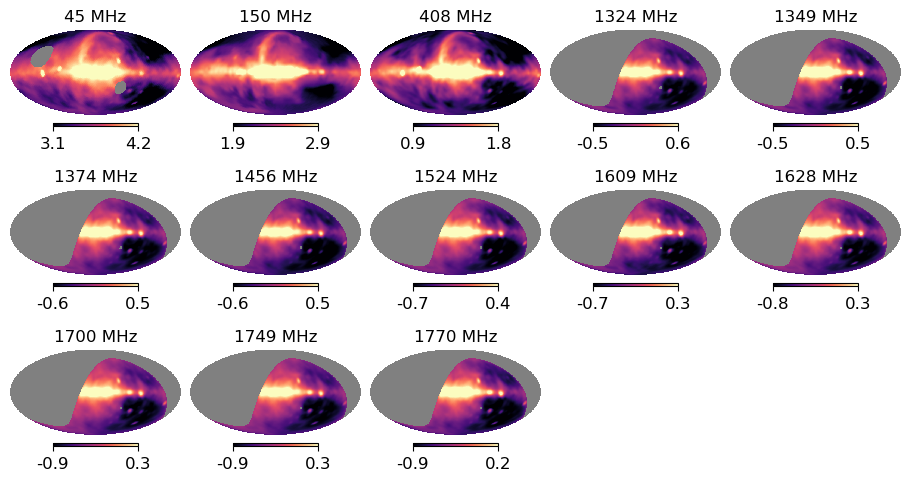}}\\
 \caption{Frequency maps at 5$^{\circ}$ resolution employed in the fit of $\beta_s$ and $c_s$ across the Southern hemisphere. Color scale is log$_{10}$(K). }
 \label{fig:radec_south}
   \end{figure*}

\begin{figure*}
 \centering
{\includegraphics[width=0.85\linewidth]{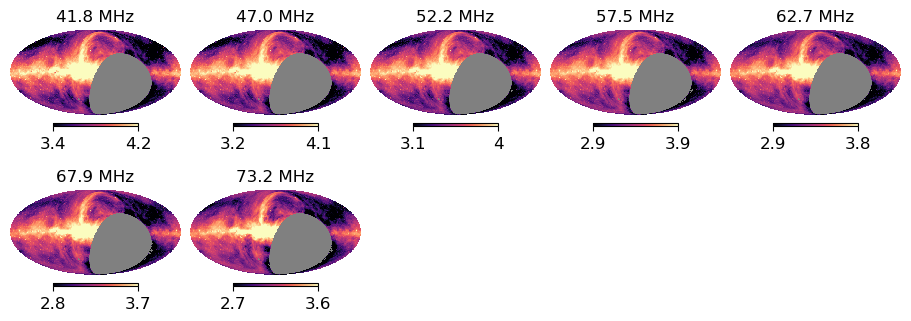}}\\
 \caption{The OVRO-LWA maps at 1$^{\circ}$ resolution employed in the fit of $\beta_s$ and $c_s$ across the Northern hemisphere. Color scale is log$_{10}$(K).}
 \label{fig:radec_ovro}
   \end{figure*}

\begin{figure*}
 \centering
{\includegraphics[width=0.85\linewidth]{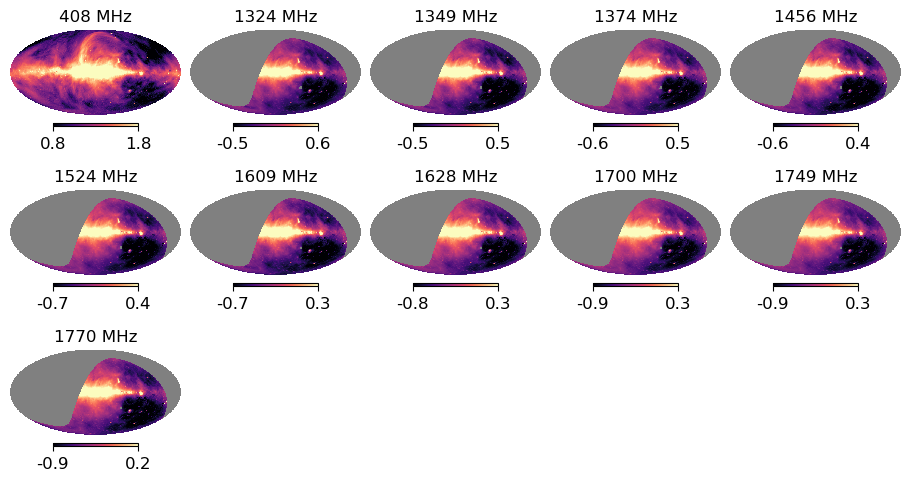}}\\
 \caption{Frequency maps at 1$^{\circ}$ resolution employed in the fit of $\beta_s$ and $c_s$ across the Southern hemisphere. Color scale is log$_{10}$(K). }
 \label{fig:radec_south_fine}
   \end{figure*}

In Fig. \ref{fig:radec}, \ref{fig:radec_ovro}, \ref{fig:radec_south}, \ref{fig:radec_south_fine}, we plot respectively the data sets employed in respectively NC, NF, SC and SF cases used to estimate   the synchrotron spectral index and the curvature parameter.

\section{Least-squares fitting}
\begin{figure*}
 \centering
  {\includegraphics[width=0.73\linewidth]{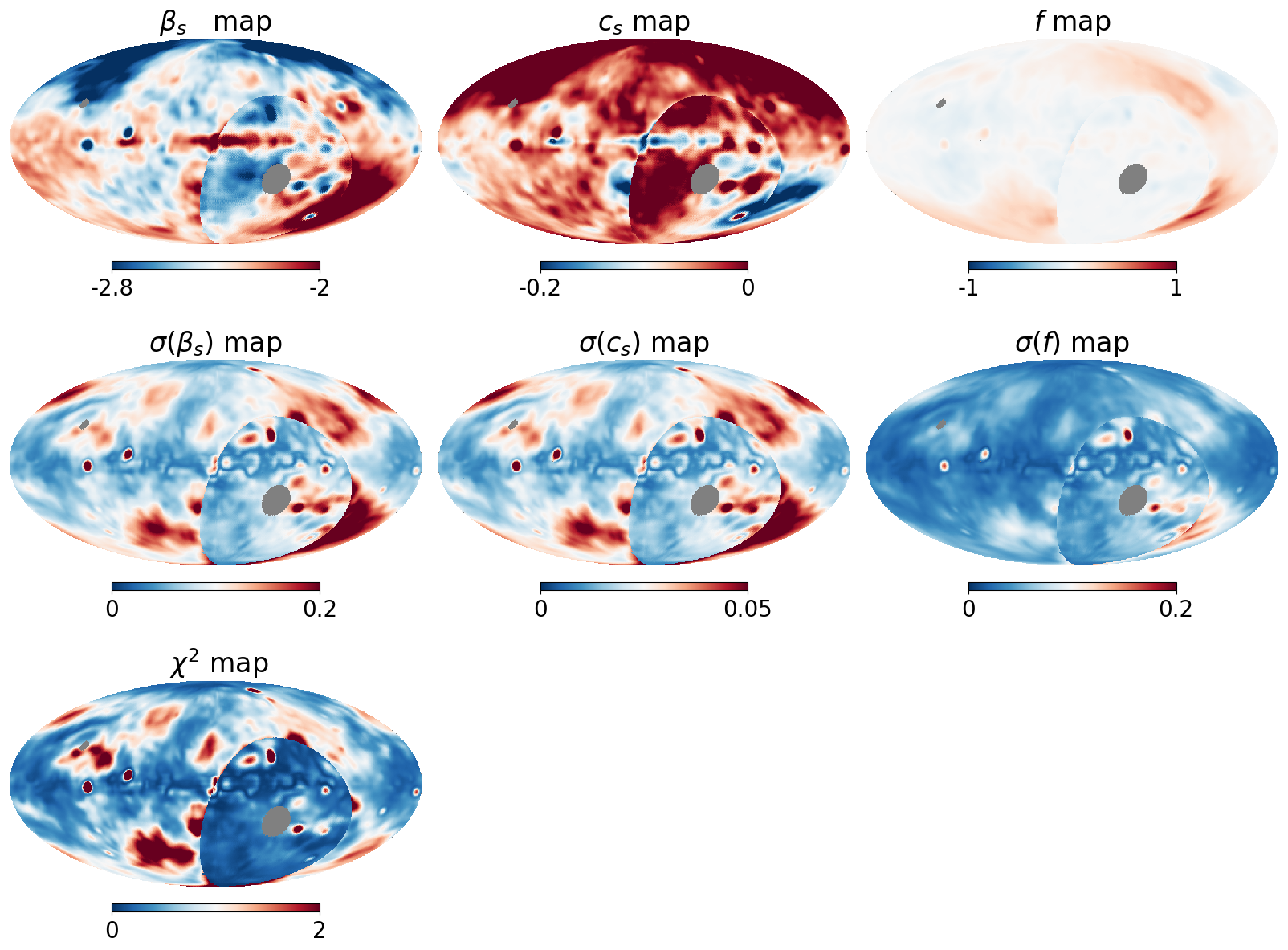}}\\
 \caption{Results of the northern and southern coarse (5$^{\circ}$) fit; the first row shows sky maps of the fitted parameters, the second presents the uncertainty maps for each of the parameters and the third row displays the per-pixel reduced $\chi^{2}$.}
 \label{fig:res5}
   \end{figure*}

\begin{figure*}
 \centering
  {\includegraphics[width=0.73\linewidth]{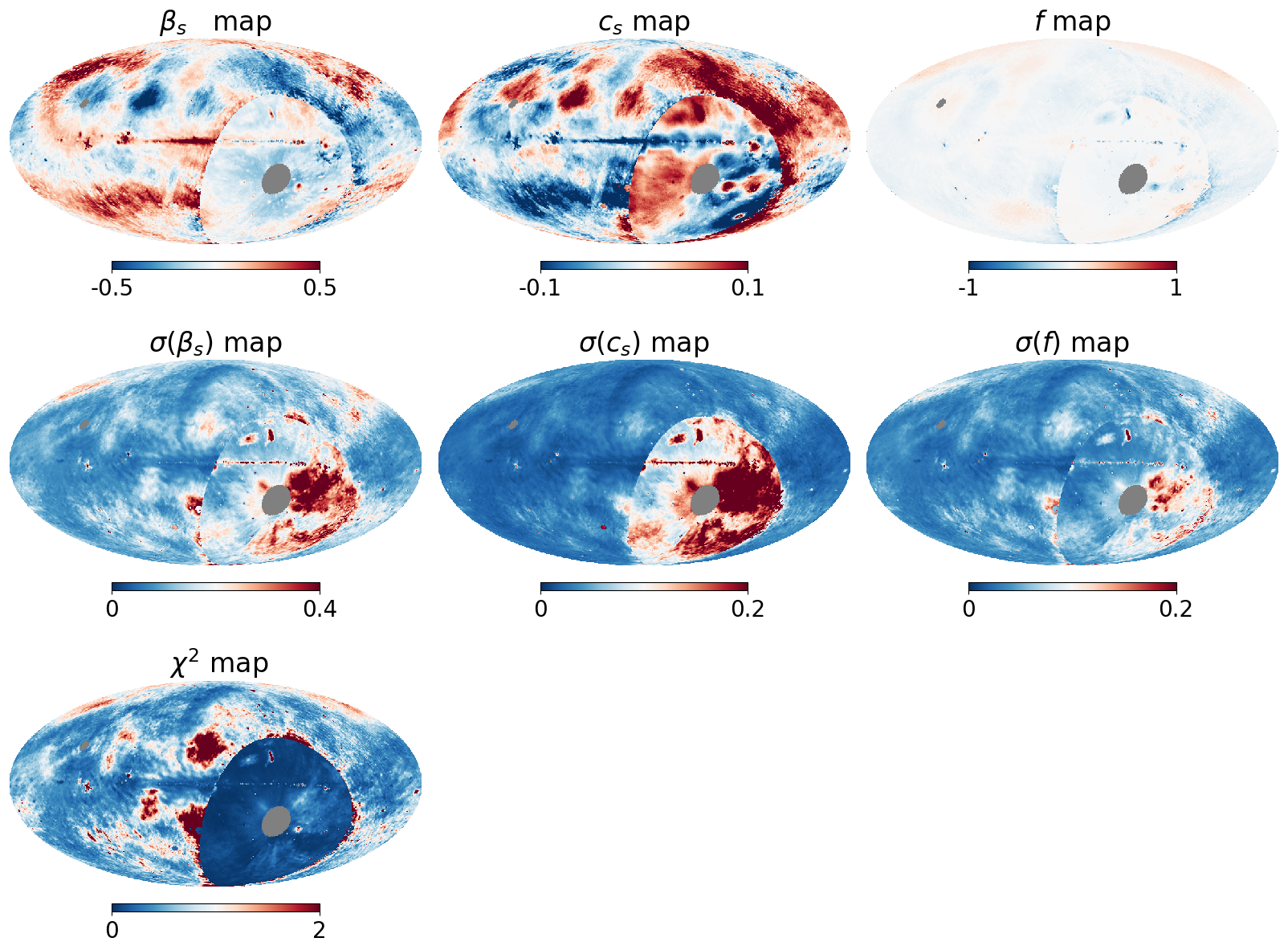}}\\
 \caption{Results of the northern and southern fine (1$^{\circ}$) fit; the first row shows sky maps of the fitted parameters, the second presents the uncertainty maps for each of the parameters and the third row displays the per-pixel reduced $\chi^{2}$.}
 \label{fig:res1}
   \end{figure*}

Figures \ref{fig:res5} and \ref{fig:res1} show results of the least-squares parametric fit for the synchrotron spectral index at 45\,MHz, curvature and $f$ amplitude parameter for each of the four fits. The northern and southern fits are presented on the same plots. 

\section{\fgb  estimates}

\begin{figure*}
 \centering  {\includegraphics[width=1\linewidth,trim=0 .2cm 0 5.cm, clip=true]{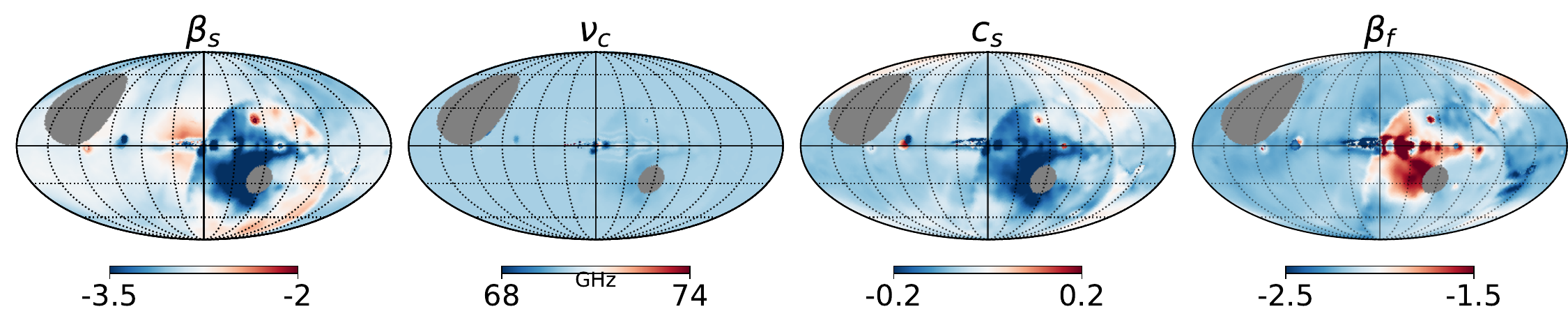}}\\
   {\includegraphics[width=1\linewidth,trim=0 .2cm 0 6.2cm, clip=true]{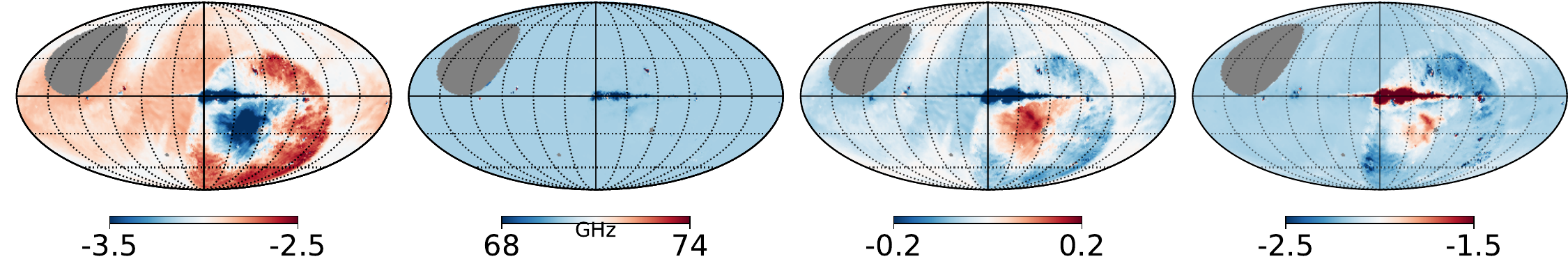}}\\
 \caption{ Spectral parameters estimated with \texttt{fgbuster}. (top) and (bottom) panels refer to the parameters estimated respectively  for the $5^{\circ}$ and $1^{\circ} $ resolution North and South maps combined into a single one. Maps  are estimated with respect to  the amplitude maps at 45 MHz and 408 MHz respectively for synchrotron and free-free emissions.} 
 \label{fig:fgbres}
   \end{figure*}

In \autoref{fig:fgbres}, we show the estimated parameters  obtained with \fgb. The NC-SC, (as well as NF-SF) have been combined together into 4 maps of the estimated parameters shown in the top (bottom) panel  of \autoref{fig:fgbres}, with unobserved pixels set to NaN values.

\begin{figure*}
 \centering  {\includegraphics[width=0.74\linewidth]{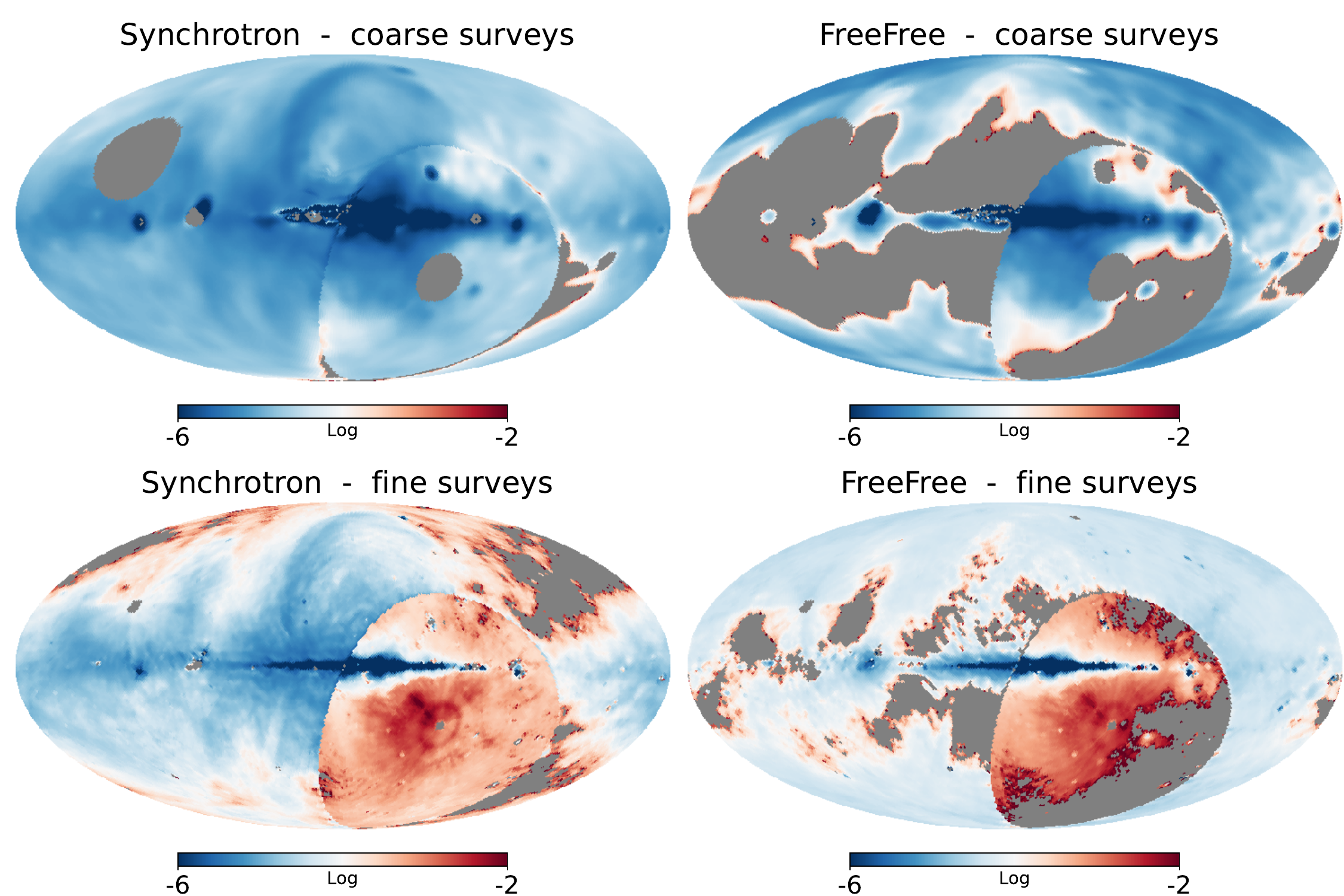}}\\
   
 \caption{ Relative uncertainties of the astrophysical components synchrotron (left) and free-free (right) separated with \fgb. \emph{Top} and \emph{bottom} panels refer to the  estimates respectively  for the $5^{\circ}$ and $1^{\circ}$ resolution, North and South maps combined into a single map. Grey regions indicate where the likelihood does not converge.}
 \label{fig:fgb_unc}

   \end{figure*}

Uncertainties with component separated maps  are estimated with respect to  the amplitude maps at 45 MHz and 408 MHz respectively for synchrotron and free-free emissions. 
 In the approach described in \citet{Stompor_2016} and summarized in \autoref{sec:fgb},
the maximum likelihood estimates obtained from \autoref{eq:like} correspond to the mean values of the spectral parameters that would be inferred from the observed data, while the curvature of the likelihood function evaluated at its maximum characterizes the level of uncertainty arising from the instrumental noise, assumed to be stationary and Gaussian. Uncertainty maps  are thus  obtained from the second derivative of the likelihood, as follows: 
\begin{displaymath}
    (\Sigma ^{-1} )_{\beta \beta'}= \left( \frac{\partial \mathcal{L}}{\partial \beta\partial \beta' } \right).
\end{displaymath}
For further details we direct the reader to \citet[Section C ] {Stompor_2016}.
We remark here that the assumption on the noise properties   leads to underestimate the  uncertainties in low SNR areas preventing convergence for the likelihood maximization. This is the reasons why we observed regions at high Galactic latitudes as well as in the outer Galactic sectors where  uncertainties of both synchrotron and free-free are  undefined.
 \fgb package combines the uncertainties onto a single uncertainty map in units of $\mu \rm{K}$. In \autoref{fig:fgb_unc}, we report the relative uncertainties on synchrotron and free-free components for both the coarse and fine resolution cases, combined together.

\bsp	
\label{lastpage}
\end{document}